**High-yield production of quantum corrals in a surface reconstruction pattern**

Wenzhen Dou[$,†,‡], Meimei Wu[$,‡], Biyu Song[†,‡], Guoxiang Zhi[‡], Chenqiang Hua[‡], Miao Zhou[*,†,‡], Tianchao Niu[*,‡]

† School of Physics, Beihang University, Beijing 100191, China

‡ Beihang Hangzhou Innovation Institute Yuhang, Hangzhou 310023, China

$ W. D. and M.W. contributed equally to this paper.

The power of surface chemistry to create atomically precise nanoarchitectures offers intriguing opportunities to advance the field of quantum technology. Strategies for building artificial electronic lattices by individually positioning atoms or molecules result in precisely tailored structures but lack structural robustness. Here, taking the advantage of strong bonding of Br atoms on noble metal surfaces, we report the production of stable quantum corrals by dehalogenation of hexabromobenzene molecules on a preheated Au(111) surface. The byproducts, Br adatoms, are confined within a new surface reconstruction pattern and aggregate into nanopores with an average size of 3.7±0.1 nm, which create atomic orbital-like quantum resonance states inside each corral due to the interference of scattered electron waves. Remarkably, the atomic orbitals can be hybridized into molecular-like orbitals with distinct bonding and anti-bonding states. Our study opens up an avenue to fabricate quantum structures with high yield and superior robustness.

Quantum confinement on the surface of noble metals has been extensively investigated since the creation of quantum corrals (QCs) by Crommie et al. in 1993 [1] with a circle of 48 Fe atoms placed on a Cu(111) surface. Atomically precise manipulation of individual atoms or molecules with low-temperature scanning tunneling microscopy (STM) enables the creation and characterization of designer quantum states of matter [2], such as hexagonal [3-4], Lieb [5], and breathing kagome lattices [6]. This approach opens up an interesting playground for artificial electronic and spin lattices with high flexibility and ability to delicately incorporate defects [7-8]. However, it is impractical for scalable applications, where a huge number of building blocks are needed for patterning into a designed architecture [9]. In contrast to atom-by-atom manipulation, self-assembly of atomically precise lattices [10] based on noncovalent intermolecular interactions takes the advantages of uniform production of supramolecular architectures with topological designation [11], constitutional tunability [12] and environmental stability over wafer-scale [13]. Regarding the robustness, covalently bonded organic structures through on-surface polymerization [14-15] hold more promise than the ones stabilized *via* noncovalent intermolecular interactions (e.g., hydrogen bonding, halogen bonding, and metal-organic coordination).

On-surface chemical reactions [16-17] have recently developed into a promising way to prepare the otherwise inaccessible architectures with desired geometry as impressively exampled by the synthesis of graphene nanoribbons with atomically precise edges and well-defined width [18-19]. However, thermally activated on-surface reactions always lead to high defect density, small domain size and limited regioselectivity [20-21]. For covalently linked organic nanopores and cycles, both entropy and in-plane strain effects coupled with competitive reaction pathways result in an ultralow yield and unwanted byproducts [22-23]. Long-range ordered two-dimensional (2D) nanoporous polymers in mesoscale lateral extension have been realized through topochemical photopolymerization [24-25], but the number of known monomers with reactive packing is very limited [26-28]. Moreover, the realization of QCs in organic nanocavities on noble metal surface requires that the pore size should be comparable to the Fermi wavelength of 2D electron gas [29-30]. These stringent requirements give rise to few experimental realizations of stable QCs in organic frameworks [31-34].

To this end, we use hexabromobenzene (HBB) and Au(111) as the precursor and substrate to construct robust QCs with high-yield production (Scheme 1). Our approach is based on the following experimental facts: First, dehalogenation reaction is appealing for on-surface synthesis of covalently linked architectures with atomic precision [18, 35-36]. The byproducts, Br adatoms, bind strongly on Au(111) surface that can only be desorbed above 720 K in the absence of other species [37]. Second, the herringbone reconstruction of Au(111) can be lifted by adsorbate-induced stress modification [38-39], which in turn makes an impact on the molecular assembly [40]. Third, Au(111) is the least reactive noble metal for Ullmann coupling that prevents the decomposition of organic skeletons before desorption [36, 41]. In this study, we reveal that hot dosing trace amount of HBB onto a preheated Au(111) surface results in QCs defined by barriers built from Br dot chain arrays. The chain arrays aggregate into highly stable QCs with an average pore size of 3.7±0.1 nm, which are confined by triangular soliton walls formed from Au(111) surface reconstruction induced by the strong bonding of Br with thermal treatment. By means of STM/STS corroborated by density functional theory (DFT) based theoretical calculations, we visualize the atomic orbital-like resonance states hosted in QCs. Remarkably, these orbitals can be hybridized into molecular-like orbitals with distinct bonding and anti-bonding states, as clearly observed in stadium-shaped QCs.

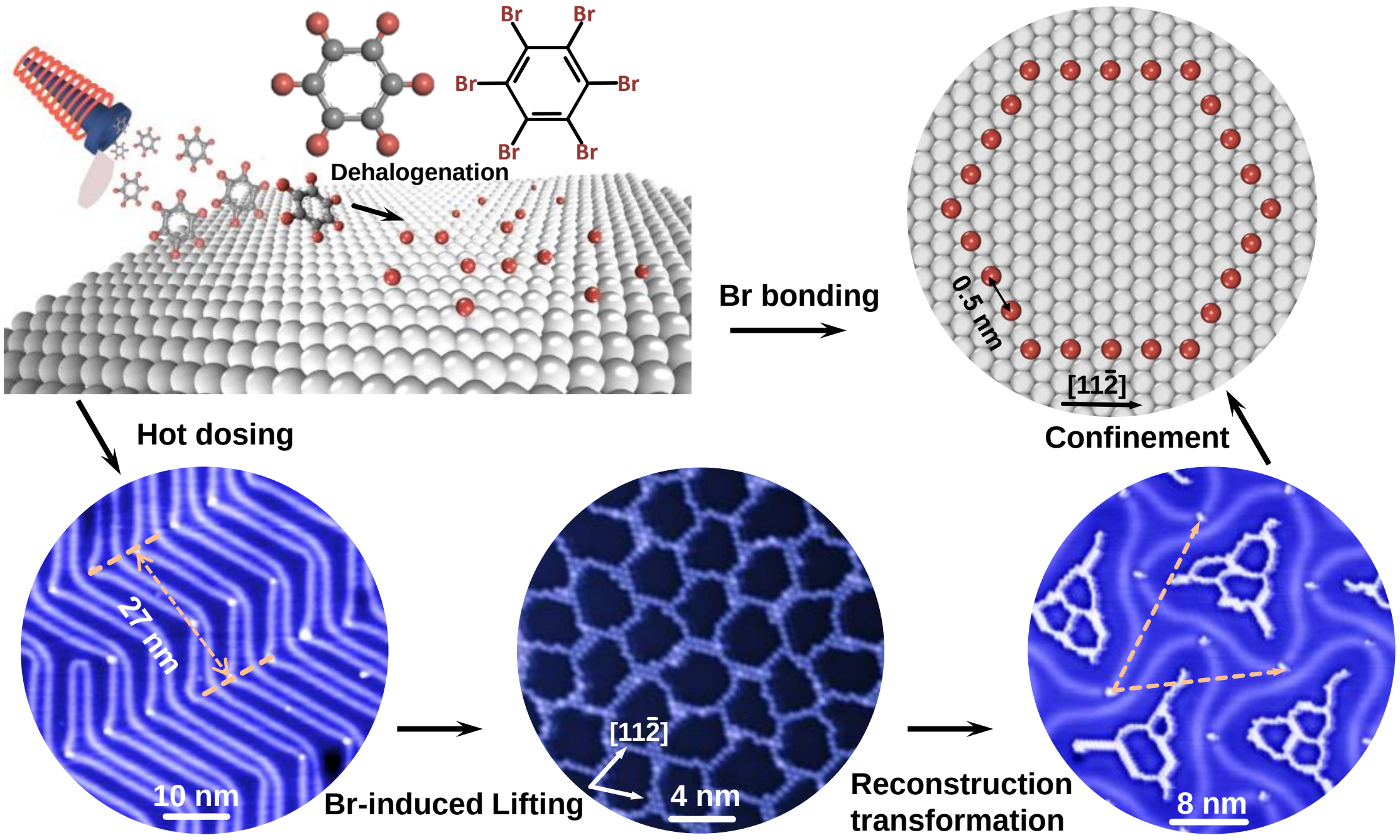

**Scheme 1.** Schematic illustration of the dehalogenation reaction of HBB and the formation of QCs on Au(111) surface. Hot dosing HBB onto a preheated Au(111) leads to the formation of web-like Br networks comprising pseudo-hexagonal Br rings. Adsorption of Br lifts the herringbone reconstruction and induces transformation to triangular soliton walls. The triangular reconstruction shows an average size of 18 nm. Br dot chain arrays are confined within the soliton walls to form QCs. One typical corral with a size matching the Fermi wavelength of Au(111) surface electrons and close to the measured average value of the pore size. Scanning parameters (left to right): -1.8 V, 100 pA; -1.2 V, 120 pA; -0.4 V, 200 pA.

Figure 1 summarizes the evolution process from linked nanoporous Br networks to QCs confined by reconstructed soliton walls upon thermal annealing under ultrahigh vacuum conditions. When HBB molecules are hot dosed onto a preheated Au(111) held at 353 K, they dehalogenate and form a porous network that is connected by linearly ordered dot arrays (Figure 1A). Each dot is assigned to an individual Br atom (see more details in Figure S1 in the Supporting Information). Meanwhile, due to the fast and irreversible dehalogenation reaction [42], decomposed carbon radicals aggregate into islands with an average height of 3.15 Å (Figure S2). Annealing at 433 K converts the network into a web-like film spanning over the whole surface (large-scale STM image in Figure S2). Figure 1B resolves a pseudo-hexagonal network separated by single-walled dot chain arrays. The propagation direction of nanopores shows registry with the Au(111) surface (hexagons in the STM image). It is noted that each pore exhibits a dark hole at the center under a positive sample bias of 0.5 V. Such a porous network comprising Br adatoms was recently observed by low-temperature STM at 5 K as a byproduct during the Ullmann reaction of brominated organic compounds on Au(111) [43-44]. It is the surface state of Au(111) mediated by adatom interaction coupled with the accumulated charge at the interface that contributes to such periodic nanoporous networks. Within the bare area, a new surface reconstruction pattern appears, showing a triangular outline with each vertex decorated by a Br dimer.

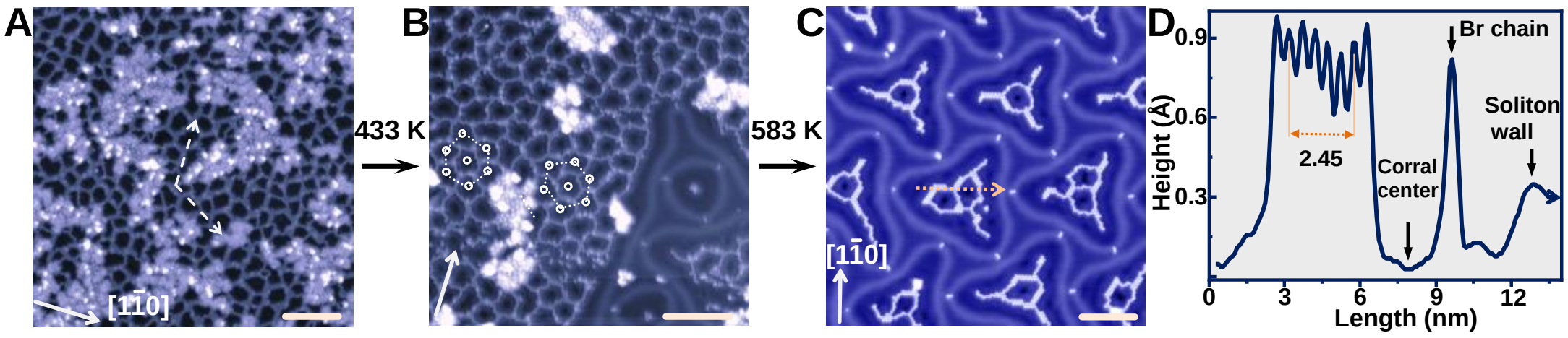

**Figure 1.** Formation of QCs from Br adatoms. (A) Hot dosing trace amount of HBB on Au(111) held at 353 K. (B) After annealing at 433 K, QCs with an average interpore distance of ~3.7±0.1 nm are formed. (C) Surface reconstruction of Au(111) with triangular soliton walls that confine Br dot arrays into QCs with pore size of ~3.7±0.1 nm upon annealing at 583 K. (D) Line profile for a QC along the arrow indicated in (C). Scanning parameters: (A) 60 nm, -1.2 V, 120 pA; (B) 50 nm, 0.5 V, 200 pA; (C) 40 nm, 0.25 V, 300 pA. Scale bar: 10 nm.

Triangular surface reconstruction pattern is obtained by annealing the sample at 583 K, which simultaneously induces a phase transition from a web-like film to isolated nanoporous Br networks (Figure 1C and Figure S2). The morphology of QCs depends crucially on the coverage of Br atoms. In particular, a large amount of Br adatoms will aggregate into densely packed hexagonal superstructures on the Au(111) surface. Half a monolayer (ML) of Br atoms lead to web-like Br networks, whereas triangular soliton walls and confined QCs are formed when the coverage is less than 1/4 ML. Within these triangular soliton walls, dotted chains are linked into porous networks and the profiles strictly follow the geometry of soliton walls. Consequently, branched mono-, binary- and ternary-pores can be observed (more images can be found in Figure S3). The typical size of nanopore is ~3.7±0.1 nm, and the average inter-pore distance is ~3.7±0.1 nm (Figure S4), close to the Fermi wavelength of Au(111) surface electrons [45]. The dot-to-dot distance is 0.5 nm, and the height of the dot is ~1 Å (line profile in Figure 1D and Figure S5). These observations reproduce well the previous studies of halogen adatoms on Au(111) surface[42]. Topography STM images of the center of these pores depend significantly on the sample bias, exhibiting a dark hole at positive sample bias, and a bright domelike spot at negative bias (Figure S6). As the size of nanopores and the typical Fermi wavelength of surface electrons are comparable, resonance states are expected to emerge due to the barriers built from Br dot chain arrays. The circular potential profile along with the strong bonding makes Br nanoporous network on Au(111) a promising candidate for robust QCs [22, 46].

Interfacial charge redistribution occurs upon the binding of Br adatoms on Au(111), giving rise to local potential variation inside the nanopores. Figure 2A shows a triangular reconstruction pattern that confines three nanopores. STS are taken at different positions to probe the local density of states (LDOS) inside each pore. As shown in Figure 2B, the dI/dV spectrum acquired at the center of a circular pore has the highest LDOS with a sharp peak at -0.13±0.02 V (labeled as $P_1$), and a shoulder peak centered at 0.42±0.1 V (labeled as $P_2$). The next highest energy level with a broad peak centered at 0.20±0.03 V is observed about 1.3 nm away from the center (labeled as $P_3$). The size and morphology of the pore affect the energy levels, as demonstrated in the dI/dV spectra taken in a stretched pore (see Figure S7), for which the $P_1$ state shifts to a higher energy of -0.19±0.02 V. It is also noted that the peak becomes broader for $P_3$ because electrons with higher energies experience lower barrier height. As a result, energy levels higher than $P_3$ cannot be distinctly resolved in the spectra. We then perform differential conductance mapping to visualize the spatial distribution of electronic states within different pores (Figure S7). Figure 2C presents the spatial distribution at different energies. Clearly, $P_1$ state of the circular pore has a domelike spot filling the pore, while $P_2$ state exhibits a bright dot at the center surrounded by a dark ring. Moreover, the dI/dV mapping acquired at 210 mV shows a dark ring-like feature, which can be assigned to $P_3$ state. STS measurements (Figure S8) and spatial mapping (Figure S9) on other pores with different topologies are also applied. It is found that the peak of the lowest energy level has the highest amplitude residing in the center of pores, and the peak position shifts to higher energies when the pore size is reduced (Figures S8 and S9).

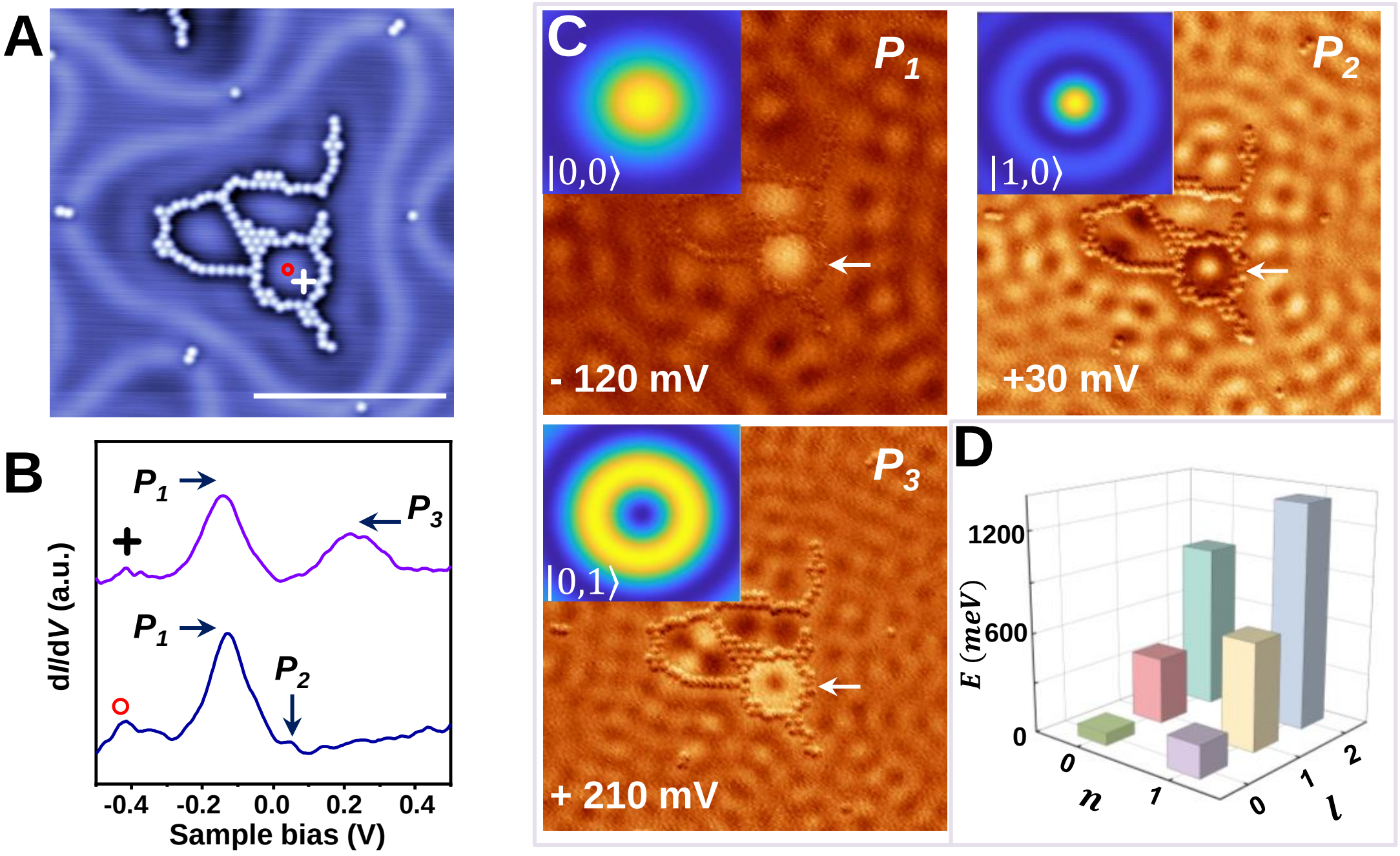

**Figure 2.** Characterization of quantum resonance states in QCs. (A) STM topography (-330 mV, 300 pA) of a triangular reconstruction with assembled Br adatoms arranged into three QCs. The Br adatoms confine the surface electrons of Au(111) into QCs. Scale bar: 10 nm. (B) Point STS acquired at different sites inside a circular pore. dI/dV curves are taken at the center (red circle) and at a point adjacent to the inner wall (white cross). (C) Constant-current dI/dV maps recorded at different energies (I = 500 pA). The insets show the theoretically calculated spatial distribution with a particle-in-box model. The energy difference of these electronic states is ~105 mV for $P_2$-$P_1$, and ~205 mV for $P_3$-$P_2$. (D) Calculated eigenvalues for different eigenstates.

The electronic states observed within these QCs can be understood by a particle-in-box model [29]. We solve the Schrödinger equation in a 2D infinite circular potential well to obtain the eigenstates and their distributions in real space (see more details in Figure S10). The calculated energy eigenvalues and eigenstates are $E_{n,l} = Z_{n,l}^2/2\mu R^2$ and $\Psi_{n,l} \propto J_l(Z_{n,l}, R)$ respectively, where $n$ is the principle quantum number, $l$ is the angular momentum quantum number, $\mu$ is the effective mass of Au(111) surface electron, $R$ is the radius of circular potential well, and $Z_{n,l}$ is the $(n+1)^{\text{th}}$ root of the $l^{\text{th}}$-order Bessel function $J_l$. As the energy eigenvalues are inversely proportional to $R^2$, smaller nanopores will have higher energy eigenvalues (Figure S10). Figure 2D shows the calculated eigenvalues for different eigenstates. The spatial distribution of the eigenstates can be determined by the particle probability $|\Psi_{n,l}|^2$. Clearly, the eigenvalues of $\Psi_{0,0}$, $\Psi_{1,0}$ and $\Psi_{0,1}$ match well with the energies of $P_1$, $P_2$ and $P_3$ states as measured by STS experiments. One-to-one correspondence between the theoretically obtained spatial distributions and the experimentally observed differential conductance mappings is also revealed, as shown in the insets of Figure 2C.

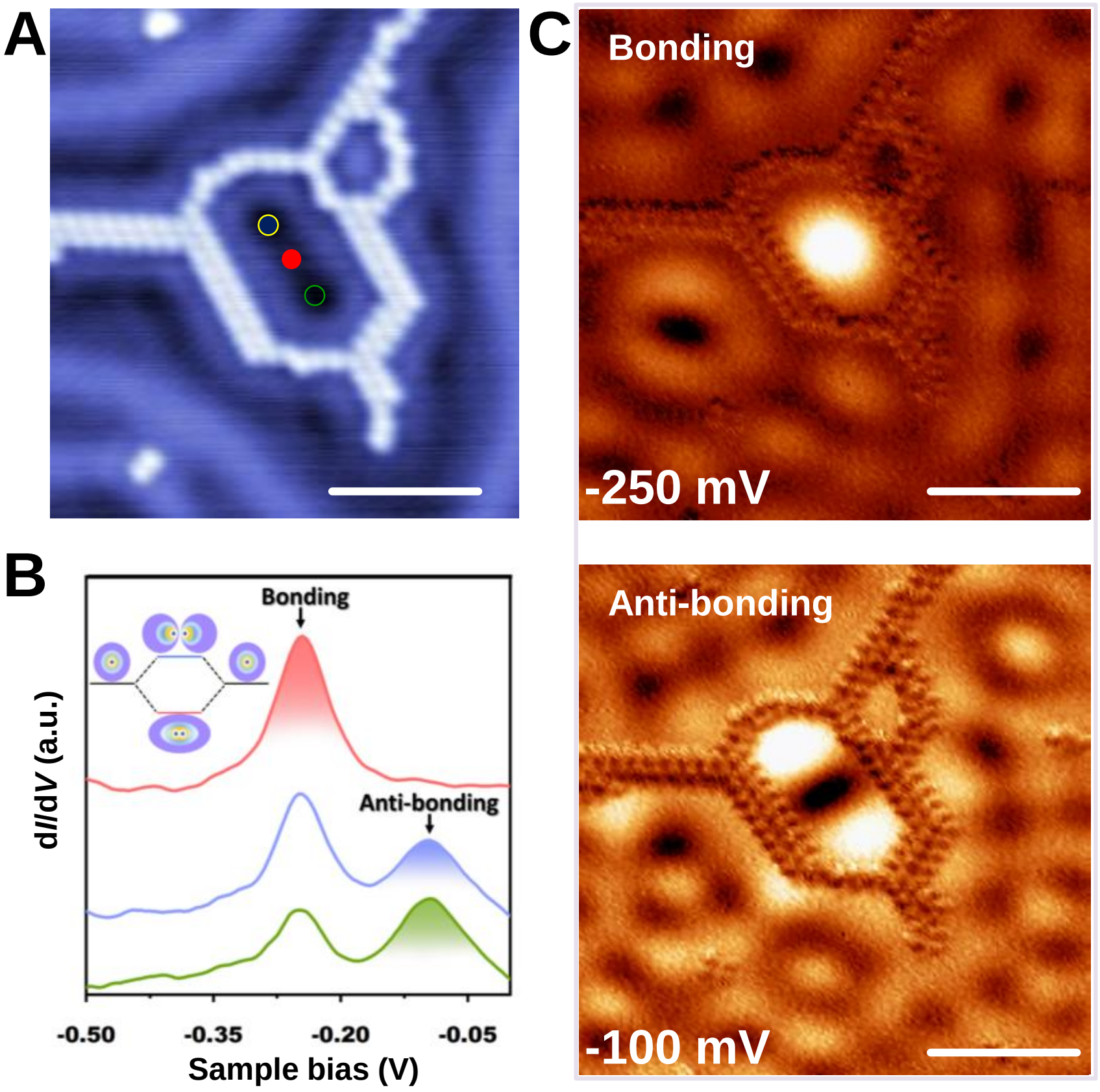

**Figure 3.** Characterization of hybridized quantum resonance states in a stadium-shaped QC. (A) Constant-current STM image (420 mV, 500 pA). (B) Point dI/dV spectra were acquired at the center (red) and two foci (blue and green) of the stadium. Inset shows the hybridization of two electronic states into bonding and anti-bonding states, similar to $H_2$-like molecular orbitals. (C) Constant-current dI/dV maps were recorded at energies of -0.25 and -0.1 V (I=500 pA), corresponding to the bonding and anti-bonding states, respectively. Scale bar: 5 nm.

The surface reconstruction induced confinement growth of nanopores also generates stadium-shaped QCs, as shown in Figure 3A. The stadium has a width of 3.7 nm with a focal length of 2.45 nm, which can be viewed as two QCs merged with no inner barrier. Within such a structure, strong coupling between the electronic states of two circular QCs can be expected that mimics orbital hybridization in homo-diatomic molecules. We measure the point STS spectra at different sites of the stadium-shaped QC (Figure S11). The spectra acquired at the center and two foci are presented in Figure 3B. Clearly, an electronic state of -0.25 V is observed predominately located at the center (red curve), while a pair of electronic states with energies of - 0.25 and -0.1 V can be found at the foci (blue and green curves). Interestingly, the energy halfway between this pair of states (-0.17 V) is close to the energy position of $P_1$ state (-0.19 V) in a circular QC. Therefore, the electronic states of this stadium-shaped QC derive from the hybridization of two $P_1$ states that split into a bonding state located at the center of the stadium and an anti-bonding state at the foci. We carry out dI/dV mapping at different energies (Figure S11), and the results at -0.25 and -0.1 V are shown in Figure 3C. Intriguingly, at -0.25 V, a domelike pattern is observed around the center of the stadium, whereas at -0.1 V, two bright dots can be seen around the two foci with a nodal line in the middle. Such peculiar spatial patterns are reminiscent of the bonding and anti-bonding states of $H_2$-like molecular orbitals (see the inset of Figure 3B, and the calculated results in Figure S12).

On Au(111) surface, uniaxial stress can be released by forming a herringbone reconstruction, in which the soliton lines connecting elbows represent domain boundaries. Since substrate relaxation is essential for stabilizing the elbows, any surface adsorbate can dramatically change the reconstruction [47]. In this work, Br adatoms assemble into linear chains to form nanopores that are confined by soliton walls, and the pore size matches well with the Fermi wavelength of Au(111). This suggests that the surface electrons in the confined soliton walls mediated by the adsorbed Br atoms provide a possible driving force for the formation of QCs.

To reveal the fundamental mechanism of forming QCs on Au(111), we perform systematic DFT based first-principles calculations on the structural, energetic and electronic properties of Br adatoms on Au(111) (see details in Methods). We first consider the adsorption of a single Br atom, which can be located at the top, bridge, hcp and fcc hollow sites of Au(111). The optimized configurations and the calculated adsorption energies are presented in Figure S13 and Table S1. It is found that Br adsorption at fcc site is the most favorable (Figure 4A). The calculated Br-Au bond length is around 2.72 Å and the adsorption energy is -1.36 eV, suggesting strong chemisorption. As Br is highly electronegative, we analyze the charge transfer between Br and Au surface. Plots of differential charge density in real space and plane-averaged charge density difference demonstrate that the charge redistribution mainly occurs at the Br-Au interface, where electrons are transferred from Au to Br (Figure 4B). Indeed, Bader charge analysis[48] shows that about 0.25 *e* is transferred from Au to Br per atom, indicative of strong ionic bonding. Interestingly, the adsorption associated with pronounced charge transfer induces noticeable structural distortion of the underlying substrate: The optimized bond length of Au-Au below Br is 3.27 Å, about 0.33 Å larger than the length in bulk Au (Figure S14), resulting in local strain on the surface. As the Br atoms propagate along the $[1\bar{1}0]$ direction of Au(111), large surface stress anisotropy can be induced[39]. Eventually, surface reconstruction transforms from herringbone to triangular soliton.

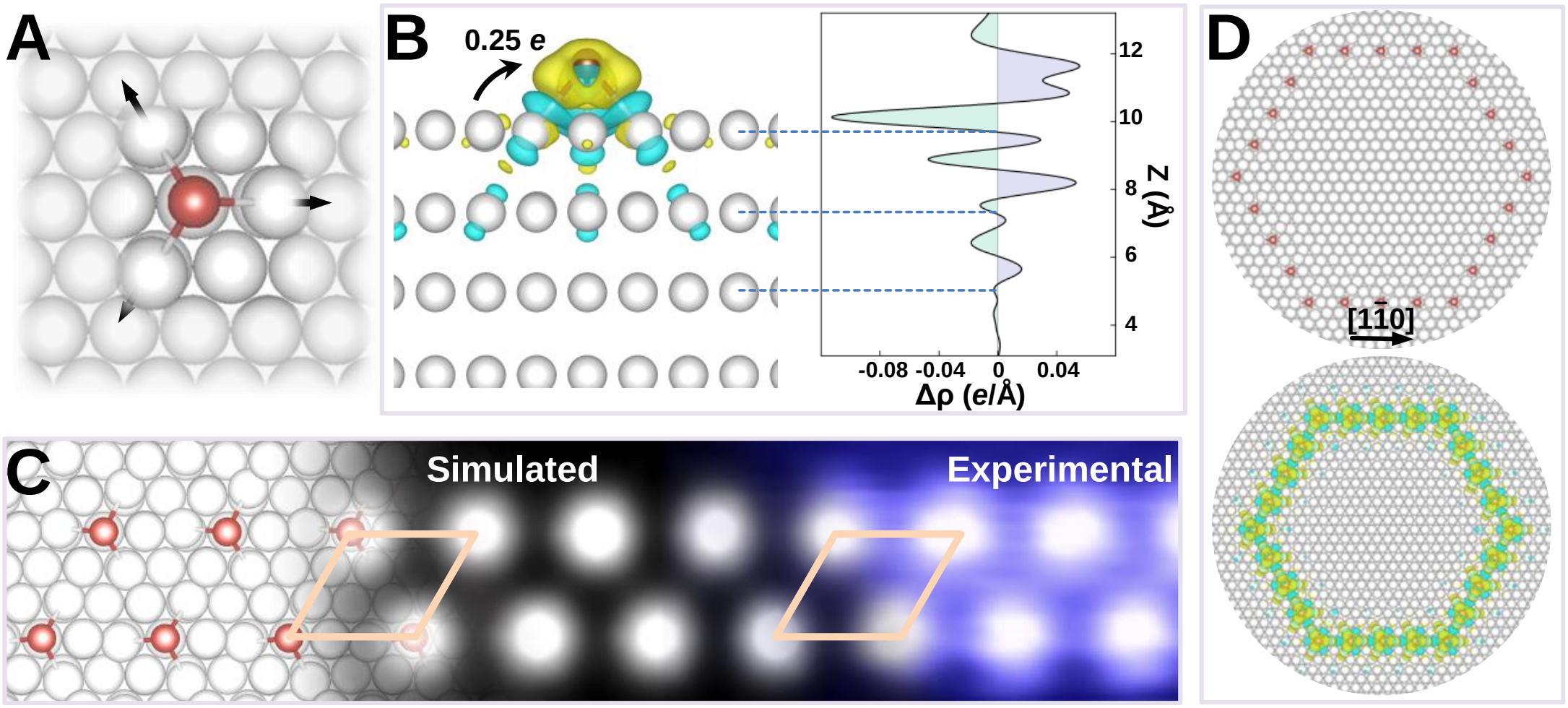

**Figure 4.** Formation of QCs by DFT calculations. (A) Optimized structure of a single Br adatom on Au(111) surface. Br atom is adsorbed at the fcc hollow site. The arrows highlight the displacement of Au atoms induced by Br. (B) Real-space charge density redistribution upon the bonding of Br on Au(111) (isovalue=0.005 $e/Å^3$). Blue and yellow colors denote electron depletion and accumulation, respectively. The plane-averaged charge density difference alone Z direction is also shown. (C) Structure of two parallel Br arrays along $[1\bar{1}0]$ direction (left), with the simulated STM image (middle) reproducing well the experimental observation (right). (D) A DFT predicted QC on Au(111), which is composed of 24 Br adatoms packing into a hexagon (top panel) with charge density redistribution shown (bottom panel).

In experiments, double chain arrays of Br on Au(111) are most frequently observed (Figures 1C and 3A). Therefore, we also build Br double chains on the surface, and the optimized structure is shown in Figure 4C. It is found that Br atoms still reside at fcc site, which aggregate into two parallel arrays with a $(\sqrt{3} \times \sqrt{3})$ pattern along $[1\bar{1}0]$ direction. The Br-Br distance is ~0.51 nm, which is out of the typical range of halogen bonds [42]. The large inter-array distance can be attributed by the strong Coulomb repulsion between negatively charged Br adatoms (Figure S15). Meanwhile, the simulated STM image of the parallel Br arrays agrees perfectly well with the experimental observation (Figure 4C). We further construct a QC defined by a hexagon with side length of five Br atoms, which involves a total of 24 Br adatoms with an interatomic distance of 0.51 nm (Figure 4D). The size of this QC is 3.5 nm (simulated STM image in Figure S16), matching well with the Fermi wavelength of Au(111) surface electrons. The calculated charge density redistribution results in a negatively charged potential profile along the hexagonal skeleton, which induces scattering of the surface electrons and thus gives rise to robust resonance states.

In this system, the strong bonding between Br adatoms and the surface induces reconstruction of Au(111) with triangular soliton walls. Following the profile of soliton walls, Br adatoms align into dot arrays that lead to the formation of QCs. It should be noted that after HBB dehalogenation, the carbon species are still adsorbed on the Au(111). These species are mobile and tend to be aggregated into larger clusters and islands (Figures S2 and S17). Although the surface reconstruction transformation and the intermediates during on-surface reaction are difficult to be captured due to the fast dehalogenation reaction process, the resulting structures and electronic states as revealed by STM/STS provide sufficient evidence for the generation of QCs. Delicate coverage control and usage of suitable halogen atoms are also crucial, as a large amount of Br adatoms on Au(111) will aggregate into densely packed hexagonal superstructures [42]. In contrast, Br adatoms on Ag(111) aggregate into short chains that are randomly distributed without forming any nanocavities (Figure S18). Moreover, unlike Br atom, Cl adsorption on Au(111) leads to the formation of surface chloride [49].

In summary, we have described an approach to create unique and robust QCs. Hot dosing HBB molecules onto a preheated Au(111) surface results in dehalogenation reaction, and the byproducts-Br adatoms-aggregate to nanopores, so that the surface electrons are confined to QCs with the profile strictly follows the surface reconstructed soliton walls. Remarkably, atomic orbital-like quantum resonance states and their hybridization into molecular-like orbitals are distinctly observed. Compared to the manipulation of individual atoms/molecules and growth of organic frameworks, the high-yield fabrication of QCs with Br nanopores strongly bonded on metal surface not only paves an avenue to explore novel quantum states of matter but also provides an exciting playground to engineer the delicately designed systems with atomic precision, a great benefit for technological applications.

**Supporting Information**

The Supporting Information is available free of charge at https://pubs.acs.org/doi/10.1021/acs.nanolett. Details of sample preparation, experimental and DFT calculation methods, large scale STM images of the QCs and carbon clusters, the size of triangular surface reconstructions and the QCs, point STS and STS mapping of various QCs, calculated results of point-in-box model, DFT results of the adsorption properties of Br and carbon species on Au(111), experimental data of Br on Ag(111).

**Corresponding Authors**

**Miao Zhou —** School of Physics, Beihang University, Beijing 100191, China; Beihang Hangzhou Innovation Institute Yuhang, Hangzhou 310023, China; http://orcid.org/0000-0003-1390-372X; Email:

mzhou@buaa.edu.cn

**Tianchao Niu** — Beihang Hangzhou Innovation Institute Yuhang, Hangzhou 310023, China; http://orcid.org/0000-0003-0502-4346; Email: niutianchao@zfau.cn

**Authors**

**Wenzhen Dou** — School of Physics, Beihang University, Beijing 100191, China; Beihang Hangzhou Innovation Institute Yuhang, Hangzhou 310023, China;

**Meimei Wu** — Beihang Hangzhou Innovation Institute Yuhang, Hangzhou 310023, China

**Biyu Song** —School of Physics, Beihang University, Beijing 100191, China; Beihang Hangzhou Innovation Institute Yuhang, Hangzhou 310023, China.

**Guoxiang Zhi** — Beihang Hangzhou Innovation Institute Yuhang, Hangzhou 310023, China

**Chenqiang Hua** — Beihang Hangzhou Innovation Institute Yuhang, Hangzhou 310023, China

**Author Contributions**

W.D. and M.W. contributed equally.

**Notes**

The authors declare no competing financial interest.

**ACKNOWLEDGMENTS**

T. Niu acknowledges the support from Dr. H. Zhu, Prof. A. Li and Prof. Z. Liu during his stay in SIMIT. This work was supported by the Natural Science Foundation of Zhejiang Province (LZ22A040004), the Science Challenge Project (TZ2018004), and the National Natural Science Foundation of China (11674042).

**REFERENCES**

1. Crommie, M. F.; Lutz, C. P.; Eigler, D. M. Confinement of Electrons to Quantum Corrals on a Metal Surface. *Science* **1993,** *262*, 218-220.
2. Khajetoorians, A. A.; Wegner, D.; Otte, A. F.; Swart, I. Creating Designer Quantum States of Matter Atom-by-Atom. *Nat. Rev. Phys.* **2019,** *1*, 703-715.
3. Gomes, K. K.; Mar, W.; Ko, W.; Guinea, F.; Manoharan, H. C. Designer Dirac Fermions and Topological Phases in Molecular Graphene. *Nature* **2012,** *483*, 306-310.
4. Zhou, M.; Ming, W.; Liu, Z.; Wang, Z.; Yao, Y.; Liu, F. Formation of Quantum Spin Hall State on Si Surface and Energy Gap Scaling with Strength of Spin Orbit Coupling. *Sci. Rep.* **2014,** *4*, 7102.
5. Slot, M. R.; Gardenier, T. S.; Jacobse, P. H.; van Miert, G. C. P.; Kempkes, S. N.; Zevenhuizen, S. J. M.; Smith, C. M.; Vanmaekelbergh, D.; Swart, I. Experimental Realization and Characterization of an Electronic Lieb Lattice. *Nat. Phys.* **2017,** *13*, 672-676.
6. Kempkes, S. N.; Slot, M. R.; van den Broeke, J. J.; Capiod, P.; Benalcazar, W. A.; Vanmaekelbergh, D.; Bercioux, D.; Swart, I.; Morais Smith, C. Robust Zero-Energy Modes in an Electronic Higher-Order Topological Insulator. *Nat. Mater.* **2019,** *18*, 1292-1297.
7. Polini, M.; Guinea, F.; Lewenstein, M.; Manoharan, H. C.; Pellegrini, V. Artificial Honeycomb

Lattices for Electrons, Atoms and Photons. *Nat. Nanotechnol.* **2013,** *8*, 625-633.
8. Yan, L.; Liljeroth, P. Engineered Electronic States in Atomically Precise Artificial Lattices and Graphene Nanoribbons. *Adv. Phys.: X* **2019,** *4*, 1651672.
9. Chen, H.; Zhang, X.-L.; Zhang, Y.-Y.; Wang, D.; Bao, D.-L.; Que, Y.; Xiao, W.; Du, S.; Ouyang, M.; Pantelides Sokrates, T.; Gao, H.-J. Atomically Precise, Custom-Design Origami Graphene Nanostructures. *Science* **2019,** *365*, 1036-1040.
10. Barth, J. V.; Costantini, G.; Kern, K. Engineering Atomic and Molecular Nanostructures at Surfaces. *Nature* **2005,** *437*, 671-679.
11. Telychko, M.; Li, G.; Mutombo, P.; Soler-Polo, D.; Peng, X.; Su, J.; Song, S.; Koh Ming, J.; Edmonds, M.; Jelínek, P.; Wu, J.; Lu, J. Ultrahigh-Yield on-Surface Synthesis and Assembly of Circumcoronene into a Chiral Electronic Kagome-Honeycomb Lattice. *Sci. Adv.* **2021,** *7*, eabf0269.
12. Dong, L.; Gao, Z. A.; Lin, N. Self-Assembly of Metal–Organic Coordination Structures on Surfaces. *Prog. Surf. Sci.* **2016,** *91*, 101-135.
13. Zhang, J. L.; Ye, X.; Gu, C.; Han, C.; Sun, S.; Wang, L.; Chen, W. Non-Covalent Interaction Controlled 2D Organic Semiconductor Films: Molecular Self-Assembly, Electronic and Optical Properties, and Electronic Devices. *Surf. Sci. Rep.* **2020,** *75*, 100481.
14. Held, P. A.; Fuchs, H.; Studer, A. Covalent-Bond Formation Via on-Surface Chemistry. *Chem. Eur. J.* **2017,** *23*, 5874-5892.
15. Clair, S.; de Oteyza, D. G. Controlling a Chemical Coupling Reaction on a Surface: Tools and Strategies for on-Surface Synthesis. *Chem. Rev.* **2019,** *119*, 4717-4776.
16. Grill, L.; Hecht, S. Covalent on-Surface Polymerization. *Nat. Chem.* **2020,** *12*, 115-130.
17. Grill, L.; Dyer, M.; Lafferentz, L.; Persson, M.; Peters, M. V.; Hecht, S. Nano-Architectures by Covalent Assembly of Molecular Building Blocks. *Nat. Nanotechnol.* **2007,** *2*, 687-691.
18. Cai, J.; Ruffieux, P.; Jaafar, R.; Bieri, M.; Braun, T.; Blankenburg, S.; Muoth, M.; Seitsonen, A. P.; Saleh, M.; Feng, X.; Müllen, K.; Fasel, R. Atomically Precise Bottom-up Fabrication of Graphene Nanoribbons. *Nature* **2010,** *466*, 470-473.
19. Wang, X.-Y.; Narita, A.; Müllen, K. Precision Synthesis Versus Bulk-Scale Fabrication of Graphenes. *Nat. Rev. Chem.* **2017,** *2*, 0100.
20. Bieri, M.; Nguyen, M.-T.; Gröning, O.; Cai, J.; Treier, M.; Aït-Mansour, K.; Ruffieux, P.; Pignedoli, C. A.; Passerone, D.; Kastler, M.; Müllen, K.; Fasel, R. Two-Dimensional Polymer Formation on Surfaces: Insight into the Roles of Precursor Mobility and Reactivity. *J. Am. Chem. Soc.* **2010,** *132*, 16669-16676.
21. Fan, Q.; Martin-Jimenez, D.; Ebeling, D.; Krug, C. K.; Brechmann, L.; Kohlmeyer, C.; Hilt, G.; Hieringer, W.; Schirmeisen, A.; Gottfried, J. M. Nanoribbons with Nonalternant Topology from Fusion of Polyazulene: Carbon Allotropes Beyond Graphene. *J. Am. Chem. Soc.* **2019,** *141*, 17713-17720.
22. Peng, X.; Mahalingam, H.; Dong, S.; Mutombo, P.; Su, J.; Telychko, M.; Song, S.; Lyu, P.; Ng, P. W.; Wu, J.; Jelínek, P.; Chi, C.; Rodin, A.; Lu, J. Visualizing Designer Quantum States in Stable Macrocycle Quantum Corrals. *Nat. Commun.* **2021,** *12*, 5895.
23. Fan, C.; Sun, B.; Li, Z.; Shi, J.; Lin, T.; Fan, J.; Shi, Z. On-Surface Synthesis of Giant Conjugated Macrocycles. *Angew. Chem. Int. Ed.* **2021,** *60*, 13896-13899.
24. Grossmann, L.; King, B. T.; Reichlmaier, S.; Hartmann, N.; Rosen, J.; Heckl, W. M.; Björk, J.; Lackinger, M. On-Surface Photopolymerization of Two-Dimensional Polymers Ordered on the Mesoscale. *Nat. Chem.* **2021,** *13*, 730-736.
25. Lackinger, M. On-Surface Synthesis – There Will Be Light. *Trends Chem.* **2022,** *4*, 471-474.

26. Palmino, F.; Loppacher, C.; Cherioux, F. Photochemistry Highlights on on-Surface Synthesis. *ChemPhysChem* **2019,** *20*, 2271-2280.
27. Lackinger, M.; Schlüter, A. D. The Current Understanding of How 2D Polymers Grow Photochemically. *Eur. J. Org. Chem* **2021,** *2021*, 5478-5490.
28. Wang, W.; Schlüter, A. D. Synthetic 2D Polymers: A Critical Perspective and a Look into the Future. *Macromol. Rapid Commun.* **2019,** *40*, 1800719.
29. Freeney, S.; Borman, S. T. P.; Harteveld, J. W.; Swart, I. Coupling Quantum Corrals to Form Artificial Molecules. *SciPost Phys.* **2020,** *9*, 085.
30. Kawai, S.; Kher-Elden, M. A.; Sadeghi, A.; Abd El-Fattah, Z. M.; Sun, K.; Izumi, S.; Minakata, S.; Takeda, Y.; Lobo-Checa, J. Near Fermi Superatom State Stabilized by Surface State Resonances in a Multiporous Molecular Network. *Nano Lett.* **2021,** *21*, 6456-6462.
31. Piquero-Zulaica, I.; Lobo-Checa, J.; Abd. EI-Fattah Z. M.; Ortega, J. E.; Klappenberger, F.; Auwärter, W.; Barth, J. V. Engineering Interfacial Quantum States and Electronic Landscapes by Molecular Nanoarchitectures. *2021, 2107.10141v1. arXiv. https://arxiv.org/abs/2107.10141 (assesed May 21, 2022).*
32. Müller, K.; Enache, M.; Stöhr, M. Confinement Properties of 2D Porous Molecular Networks on Metal Surfaces. *J. Phys. Condens. Matter* **2016,** *28*, 153003.
33. Chen, M.; Shang, J.; Wang, Y.; Wu, K.; Kuttner, J.; Hilt, G.; Hieringer, W.; Gottfried, J. M. On-Surface Synthesis and Characterization of Honeycombene Oligophenylene Macrocycles. *ACS Nano* **2017,** *11*, 134-143.
34. Shu, C.-H.; He, Y.; Zhang, R.-X.; Chen, J.-L.; Wang, A.; Liu, P.-N. Atomic-Scale Visualization of Stepwise Growth Mechanism of Metal-Alkynyl Networks on Surfaces. *J. Am. Chem. Soc.* **2020,** *142*, 16579-16586.
35. Galeotti, G.; De Marchi, F.; Hamzehpoor, E.; MacLean, O.; Rajeswara Rao, M.; Chen, Y.; Besteiro, L. V.; Dettmann, D.; Ferrari, L.; Frezza, F.; Sheverdyaeva, P. M.; Liu, R.; Kundu, A. K.; Moras, P.; Ebrahimi, M.; Gallagher, M. C.; Rosei, F.; Perepichka, D. F.; Contini, G. Synthesis of Mesoscale Ordered Two-Dimensional π-Conjugated Polymers with Semiconducting Properties. *Nat. Mater.* **2020,** *19*, 874-880.
36. Gao, W.; Kang, F.; Qiu, X.; Yi, Z.; Shang, L.; Liu, M.; Qiu, X.; Luo, Y.; Xu, W. On-Surface Debromination of $C_6Br_6$: $C_6$ Ring Versus $C_6$ Chain. *ACS Nano* **2022**, *16*, 6578-6584.
37. Mairena, A.; Baljozovic, M.; Kawecki, M.; Grenader, K.; Wienke, M.; Martin, K.; Bernard, L.; Avarvari, N.; Terfort, A.; Ernst, K. H.; Wackerlin, C. The Fate of Bromine after Temperature-Induced Dehydrogenation of on-Surface Synthesized Bisheptahelicene. *Chem. Sci.* **2019,** *10*, 2998-3004.
38. Maksymovych, P.; Sorescu, D. C.; Yates, J. T. Gold-Adatom-Mediated Bonding in Self-Assembled Short-Chain Alkanethiolate Species on the Au(111) Surface. *Phys. Rev. Lett.* **2006,** *97*, 146103.
39. Sun, J. T.; Gao, L.; He, X. B.; Cheng, Z. H.; Deng, Z. T.; Lin, X.; Hu, H.; Du, S. X.; Liu, F.; Gao, H. J. Surface Reconstruction Transition of Metals Induced by Molecular Adsorption. *Phys. Rev. B* **2011,** *83*, 115419.
40. Inayeh, A.; Groome, R. R. K.; Singh, I.; Veinot, A. J.; de Lima, F. C.; Miwa, R. H.; Crudden, C. M.; McLean, A. B. Self-Assembly of N-Heterocyclic Carbenes on Au(111). *Nat. Commun.* **2021,** *12*, 4034.
41. Gao, H.-Y.; Held, P. A.; Knor, M.; Mück-Lichtenfeld, C.; Neugebauer, J.; Studer, A.; Fuchs, H. Decarboxylative Polymerization of 2,6-Naphthalenedicarboxylic Acid at Surfaces. *J. Am. Chem. Soc.* **2014,** *136*, 9658-9663.
42. Niu, T.; Wu, J.; Ling, F.; Jin, S.; Lu, G.; Zhou, M. Halogen-Adatom Mediated Phase Transition of

Two-Dimensional Molecular Self-Assembly on a Metal Surface. *Langmuir* **2018,** *34*, 553-560.
43. Kawai, S.; Krejčí, O.; Nishiuchi, T.; Sahara, K.; Kodama, T.; Pawlak, R.; Meyer, E.; Kubo, T.; Foster Adam, S. Three-Dimensional Graphene Nanoribbons as a Framework for Molecular Assembly and Local Probe Chemistry. *Sci. Adv.* **2020,** *6*, eaay8913.
44. Zhong, Q.; Niu, K.; Chen, L.; Zhang, H.; Ebeling, D.; Björk, J.; Müllen, K.; Schirmeisen, A.; Chi, L. Substrate-Modulated Synthesis of Metal–Organic Hybrids by Tunable Multiple Aryl–Metal Bonds. *J. Am. Chem. Soc.* **2022,** *144*, 8214-8222.
45. Sotthewes, K.; Nijmeijer, M.; Zandvliet, H. J. W. Confined Friedel Oscillations on Au(111) Terraces Probed by Thermovoltage Scanning Tunneling Microscopy. *Phys. Rev. B* **2021,** *103*, 245311.
46. Piquero-Zulaica, I.; Lobo-Checa, J.; Sadeghi, A.; El-Fattah, Z. M. A.; Mitsui, C.; Okamoto, T.; Pawlak, R.; Meier, T.; Arnau, A.; Ortega, J. E.; Takeya, J.; Goedecker, S.; Meyer, E.; Kawai, S. Precise Engineering of Quantum Dot Array Coupling through Their Barrier Widths. *Nat. Commun.* **2017,** *8*, 787.
47. Jewell, A. D.; Tierney, H. L.; Sykes, E. C. H. Gently Lifting Gold's Herringbone Reconstruction: Trimethylphosphine on Au(111). *Phys. Rev. B* **2010,** *82*, 205401.
48. Bader, R. F. W., *Atoms in Molecules : A Quantum Theory*. Oxford : Clarendon press: 1990.
49. Gao, W.; Baker, T. A.; Zhou, L.; Pinnaduwage, D. S.; Kaxiras, E.; Friend, C. M. Chlorine Adsorption on Au(111): Chlorine Overlayer or Surface Chloride? *J. Am. Chem. Soc.* **2008,** *130*, 3560-3565.

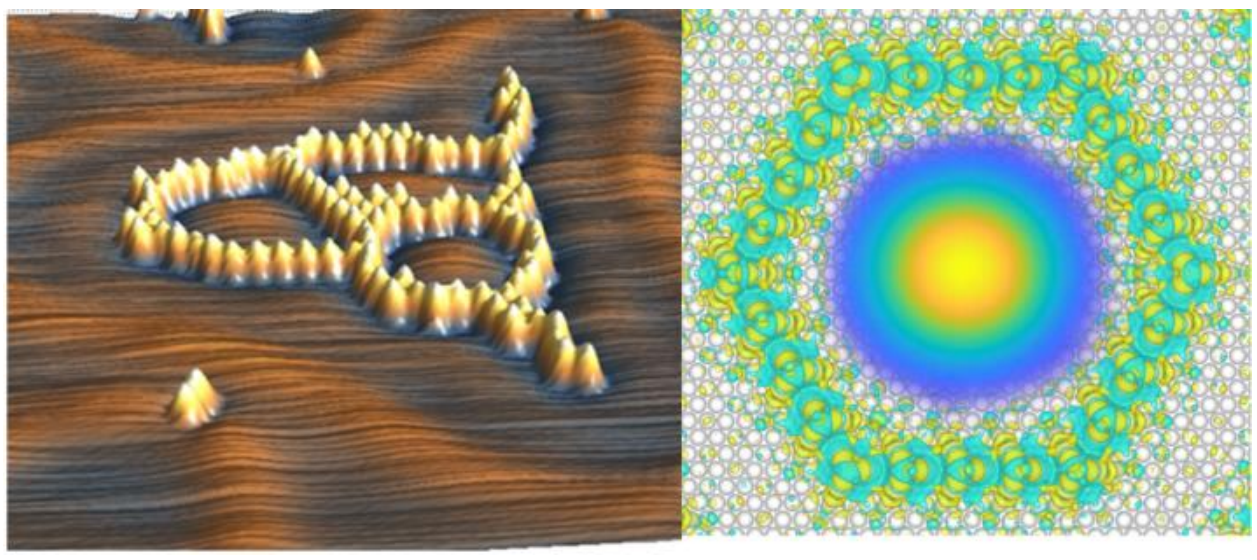

Table of Contents

Supporting Information For

# High-yield production of quantum corrals in a surface reconstruction pattern

Wenzhen Dou[$,†,‡], Meimei Wu[$,‡], Biyu Song[†,‡], Guoxiang Zhi[‡], Chenqiang Hua[‡], Miao Zhou[*,†,‡], Tianchao Niu[*,‡]

† School of Physics, Besihang University, Beijing 100191, China

‡ Beihang Hangzhou Innovation Institute Yuhang, Hangzhou 310023, China

$ W.D. and M.W. contributed equally to this paper.

* E-mail: M.Z. (mzhou@buaa.edu.cn); T.N. (niutianchao@zfau.cn)

## EXPERIMENTAL METHODS

**Sample preparation**. Atomically clean Au(111) surfaces were prepared through repeated cycles of $Ar^+$ bombardment (1 kV, $1\times10^{-5}$ mbar) and annealing (700 K, 15 min). To delicately control the coverage of Br, only a trace amount of HBB was sublimated from a Knudsen cell at crucible temperature of 80 °C onto Au(111) surface held at different substrate temperatures.

**STM measurements**. All STM experiments were carried out using a commercial SPECS JT-SPM held at T=4.5 K with a chemically etched tungsten tip. STS and differential conductance mapping were performed using a lock-in amplifier with a modulation frequency of 617 Hz and amplitudes of 4 to 20 mV. STM images and STS maps were processed by WSXM software [1].

**First-principles calculations.** First-principles calculations were performed within the Vienna ab initio simulation package (VASP) using the projector augmented-wave (PAW) approach [2-3]. The wave functions were expanded in a plane-wave basis set with a kinetic energy cutoff of 400 eV. The electronic exchange-correlation energy was described with the Perdew-Burke-Ernzerhof (PBE) formalism of the generalized gradient approximation (GGA) [4], and the semi-empirical Grimme's scheme (DFT-D3) [5] was adopted for dispersion correction. Au(111) surface was modeled by using a 2×3 tetragonal supercell with five atomic layers, and a Γ-centered 4×3×1 Monkhorst–Pack *k*-point mesh was sampled in the first Brillouin zone. A vacuum region of over 15 Å in the direction perpendicular to the slab was used to eliminate spurious interactions between periodic images. For the adsorption of Br atoms, the bottom three Au layers were fixed while all other atoms were fully relaxed until the total energy and atomic forces were smaller than $10^{-5}$ eV and 0.01 eV/Å, respectively. STM images were simulated within the Tersoff-Hamann approximation [6]. The adsorption energy of Br, $E_{ad}$, was calculated by, $E_{ad} = E_{total} - E_{Au} - \frac{1}{2}E_{Br_2}$, where $E_{total}$, $E_{Au}$ and $E_{Br_2}$ are the energies of the adsorbed system, pure Au (111) surface, and $Br_2$ in the gas phase, respectively. The differential charge density, $\Delta\rho$, was calculated by, $\Delta\rho = \rho_{total} - \rho_{Au} - \rho_{Br}$, with $\rho_{total}$, $\rho_{Au}$, and $\rho_{Br}$ representing charge densities of the adsorbed system, Au(111) surface and adsorbed Br atom, respectively.

## EXPERIMENTAL AND THEORETICAL RESULTS

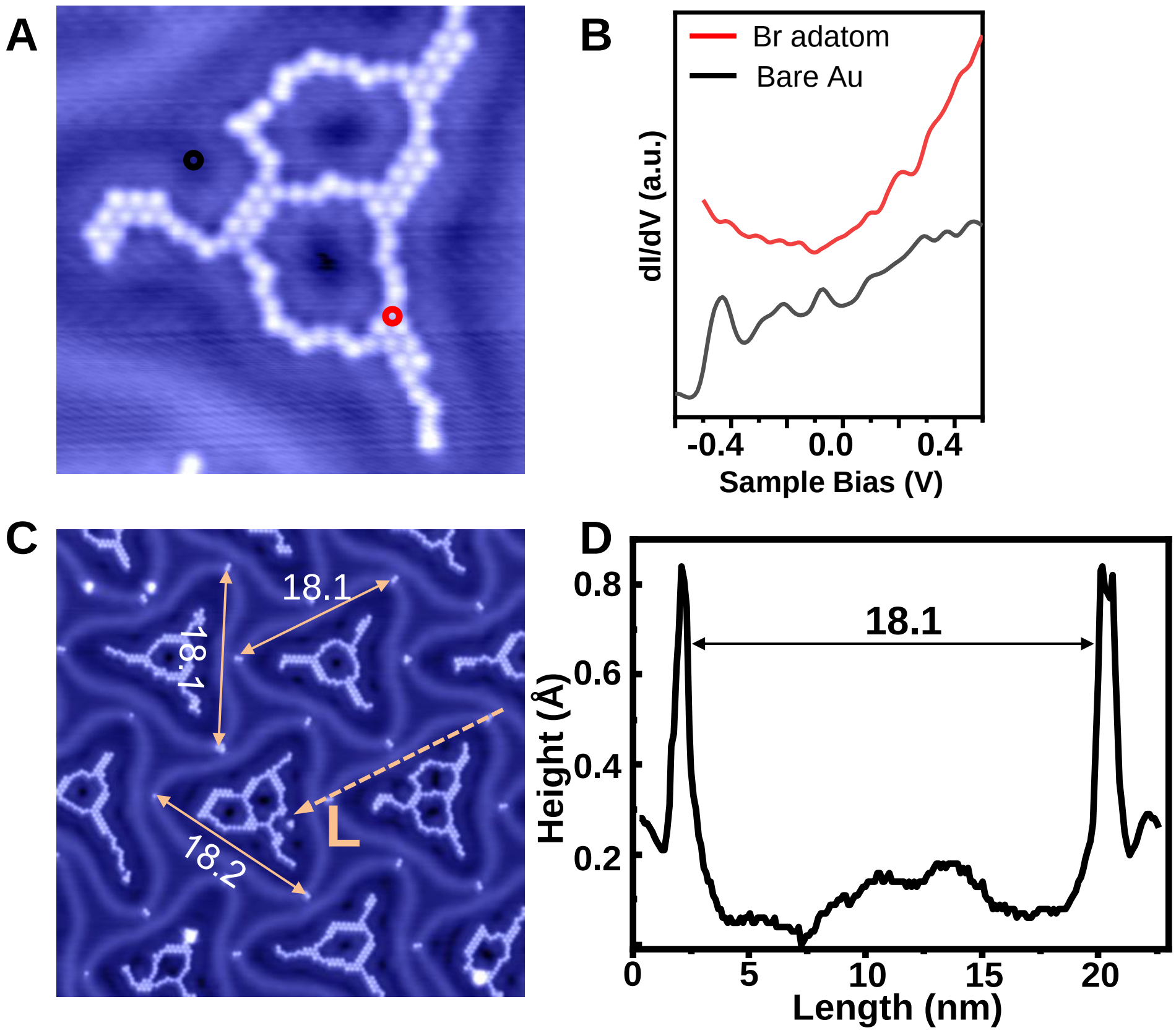

**Figure S1**. (A) STM image of Br adatoms aggregated into a nanoporous network after annealing. (B) Differential conductance spectra acquired over the Br adatom and bare Au surface. (C) STM image of the triangular surface reconstruction of Au(111) with an average size of 18 nm. (D) Line profile measured along L as indicated in (A). Scanning parameters: (A) 20 nm, 250 mV, 40 pA; (C) 40 nm, 250 mV, 300 pA.

Figure S1B shows the STS spectra acquired over aggregated dot arrays, which were compared to the spectra acquired over bare Au surface. The Au(111) surface-state band edge was observed at around -0.4 V (black line). The signature for identifying the bare Au surface was not observed over the dot arrays (red line). This is consistent with previous reports on Br [7], which quenches the surface-state band edge of noble metals. Therefore, each bright dot can be assigned to an individual Br atom.

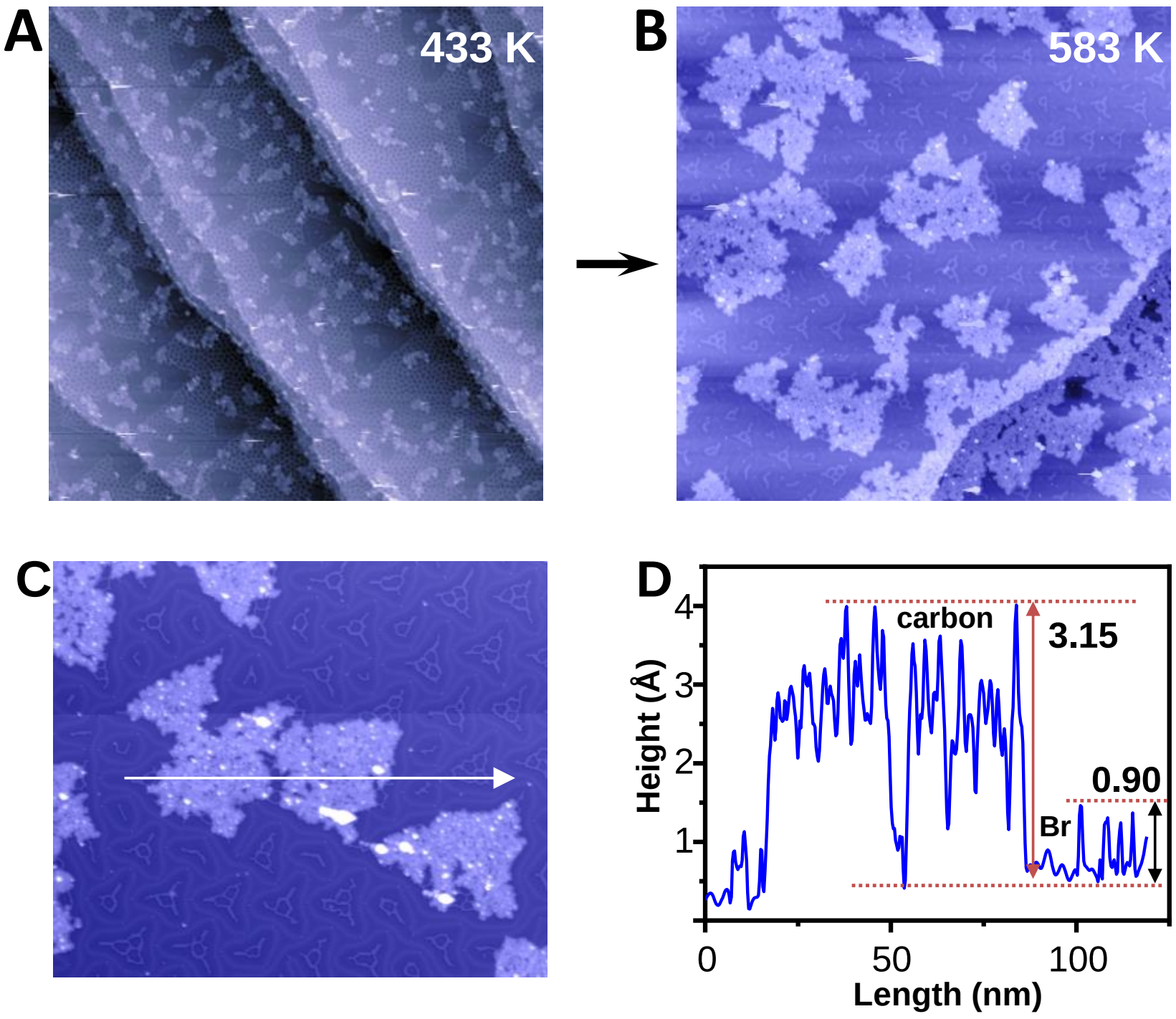

**Figure S2.** STM images showing the formation of QCs *via* dehalogenation of HBB on Au(111). (A) Annealing at 433 K leads to web-like films spanning over the whole surface. (B) QCs defined by triangular reconstruction are formed after annealing at 583 K. (C) STM image of carbon clusters and QCs. (D) Line profile along the arrow indicated in (C), showing the height of carbon species and Br adatoms. Scanning parameters: (A) 350 nm, 1.2 V, 80 pA; (B) 250 nm, 1.2 V, 88 pA. (C) 150 nm, -0.5 V, 100 pA.

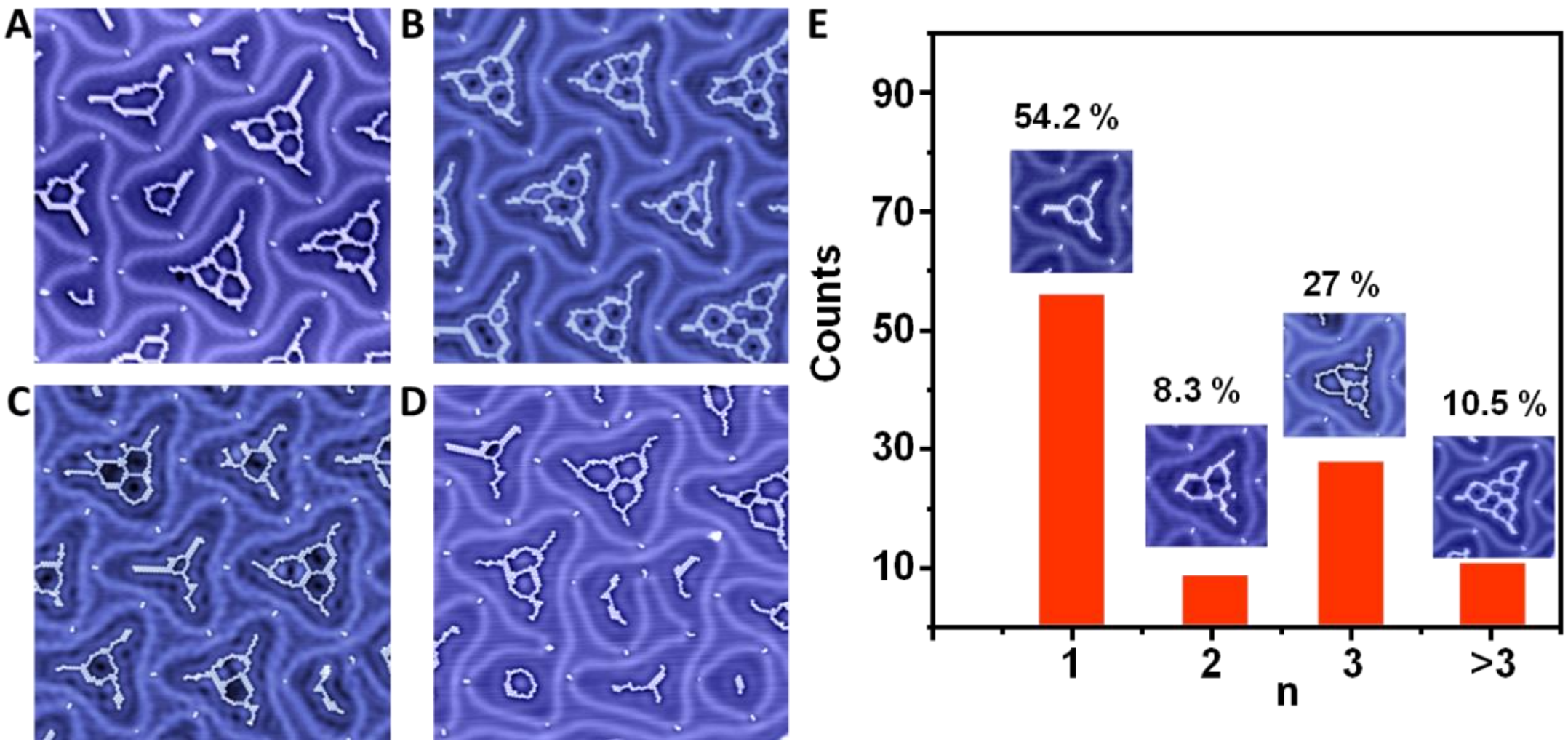

**Figure S3.** (A-D) Representative STM images showing QCs with different topologies. The profiles of QCs strictly follow the geometry of soliton walls. (E) Statistic analyses of QCs with different number of nanopores. Scanning parameters: (A) 50 nm, -500 mV, 500 pA; (B) 50 nm, 300 mV, 200 pA; (C) 50 nm, 600 mV, 100 pA; (D) 50 nm, -330 mV, 300 pA.

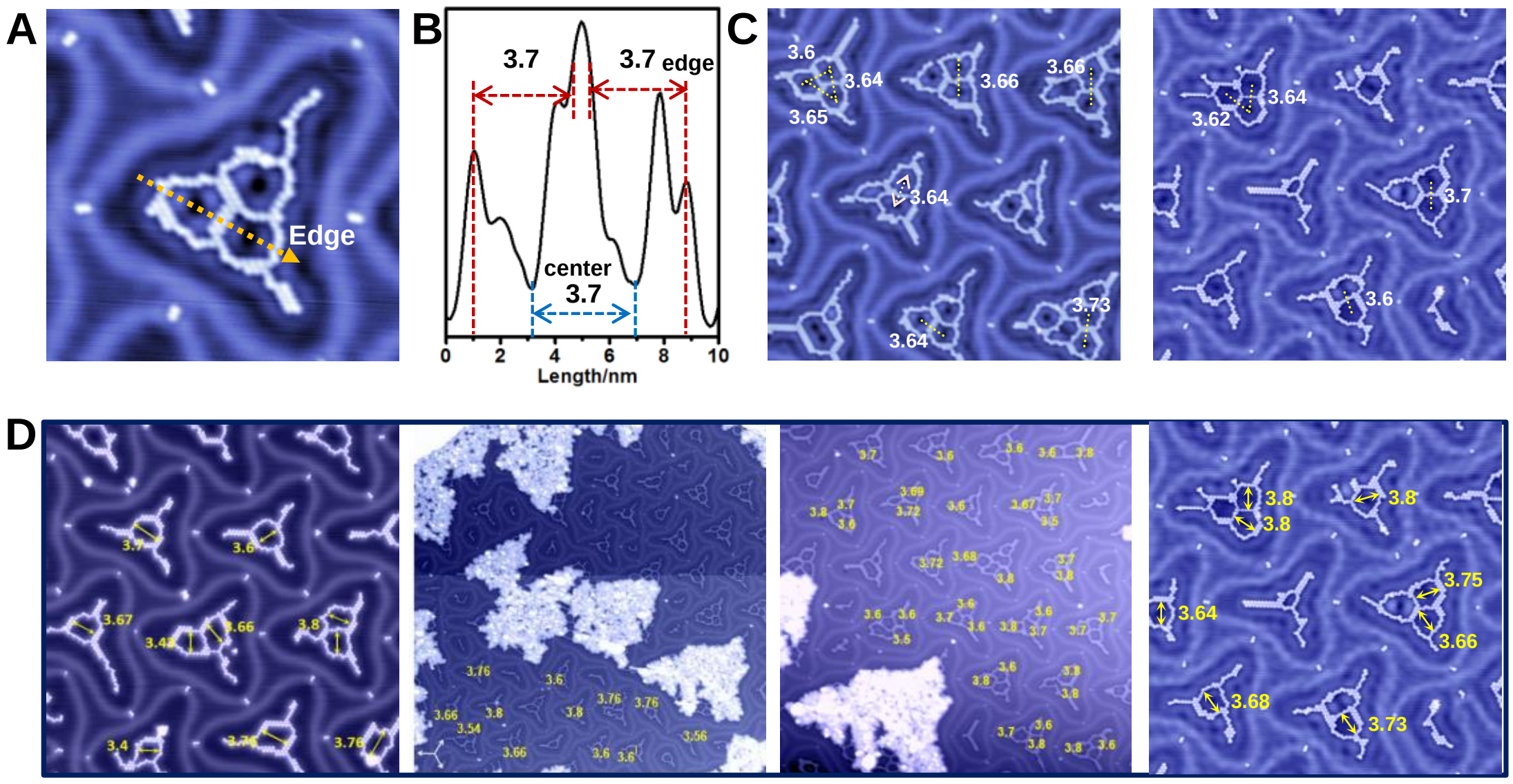

**Figure S4**. Measurements on the average size and interpore distance of QCs. (A) STM image of three nanopores confined within one reconstructed triangle. (B) A line profile taken across two nanopores along the arrow indicated in (A). (C) The average interpore distance measured to be ~3.7±0.1 nm, close to the Fermi wavelength of Au(111) surface electron. (D) STM images of various QCs. Scanning parameters: (A) 22 nm, 100 mV, 300 pA; (C) 50 nm, 500 mV, 500 pA; 50 nm, 100 mV, 100 pA. (D) From left to right: 50 nm, 400 mV, 300 pA; 150 nm, -500 mV, 100 pA; 100 nm, -400 mV, 200 pA; 50 nm, 100 mV, 100 pA.

We have measured the size of a considerable number of QCs. As shown in Figure S4, the size mainly falls within the range of 3.6~3.8 nm. We define the average pore size and interpore distance to be 3.7±0.1 nm.

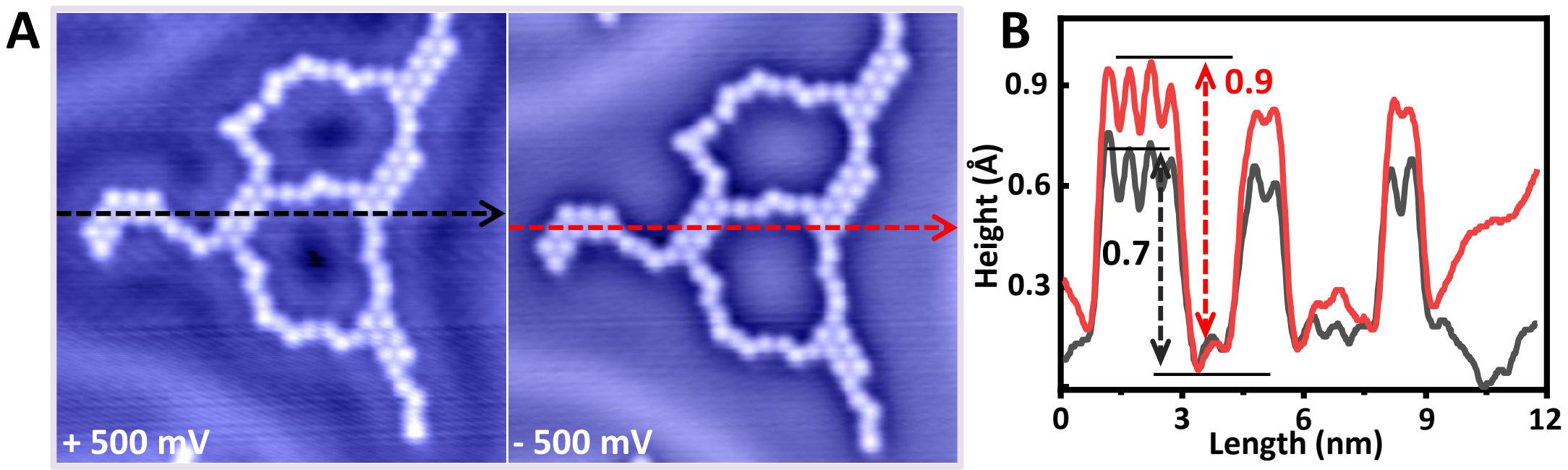

**Figure S5.** Bias dependent STM images showing the height difference of Br chain arrays. (A) At positive bias, the QC shows a dark hole at center. (B) The QC exhibits a domelike spot at negative bias. The measured height difference is ~0.2 Å. Scanning parameters: 12 nm, 500 pA.

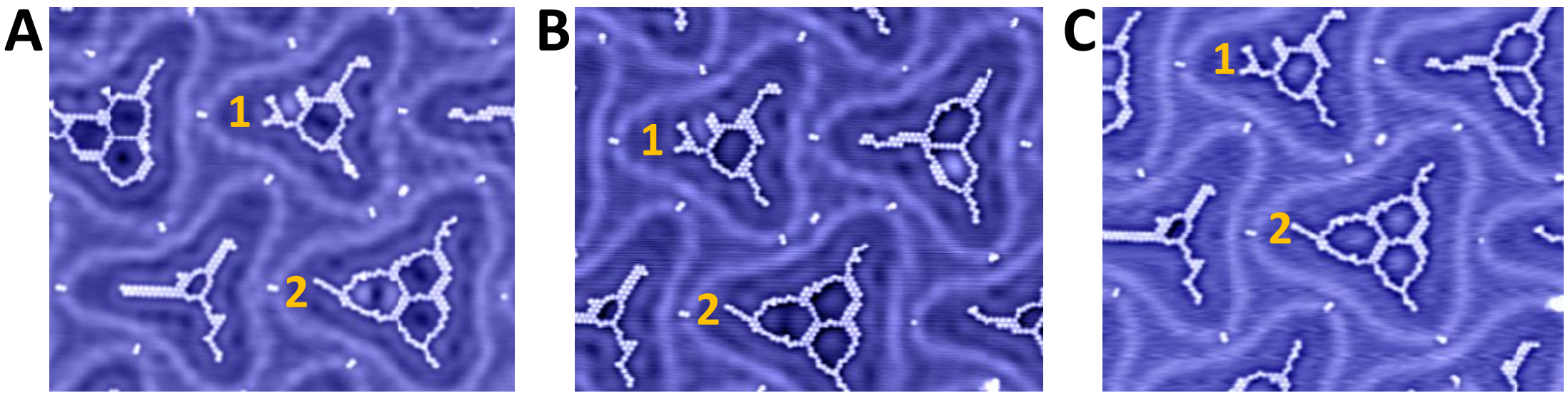

**Figure S6**. Bias dependent features of QCs. The center of QC appears a dark hole at positive bias, and a bright domelike spot at negative bias. The numbers 1 and 2 are used to guide the eyes, showing the position of measured QCs in different images. Scanning parameters: 40×33 nm, (A) 100 mV, 100 pA; (B) -100 mV, 100 pA; (C) -330 mV, 300 pA.

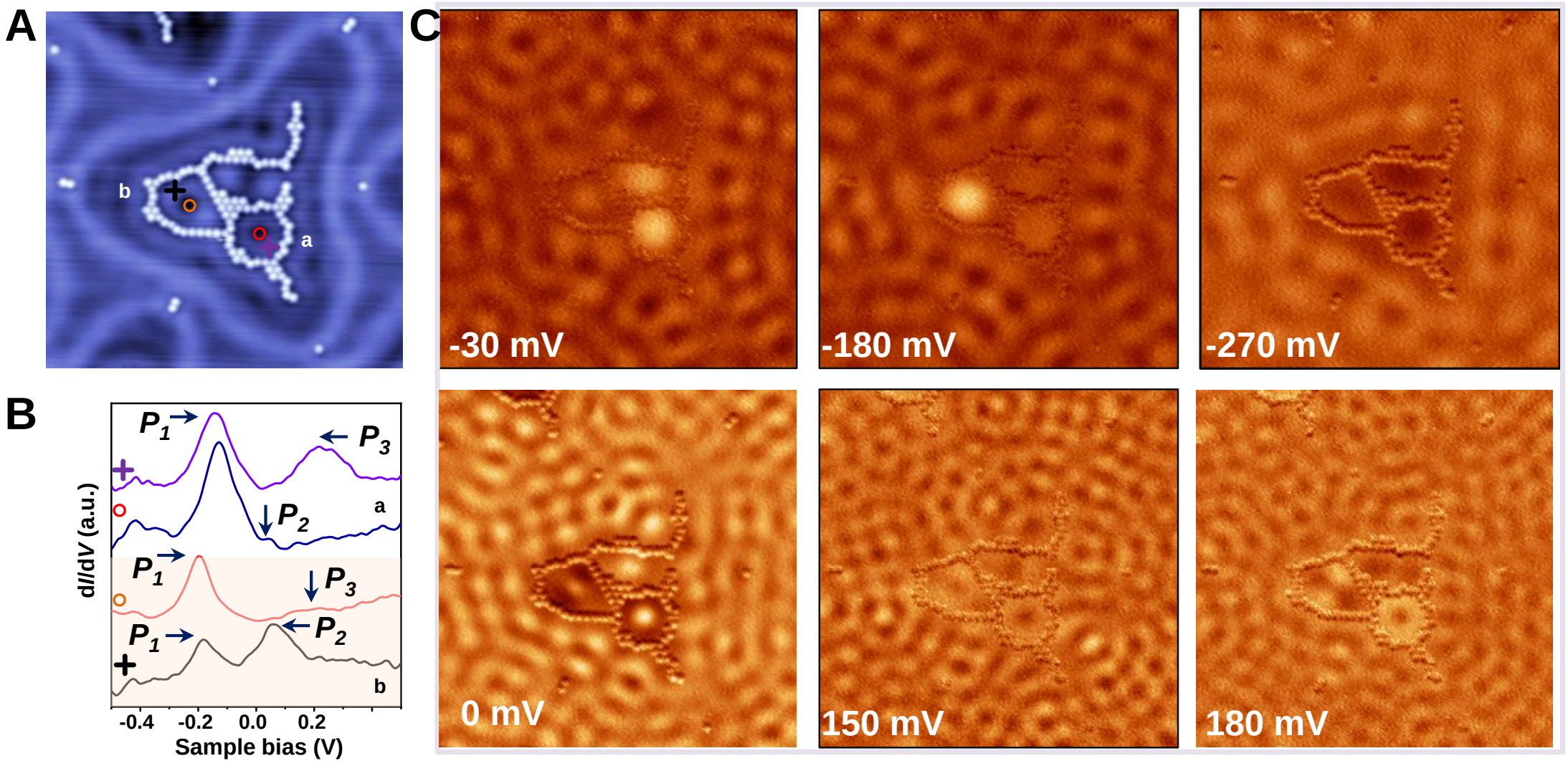

**Figure S7.** (A) STM image of three QCs confined in one triangle. (B) Point STS taken at different areas marked by colored circles and crosses in panel (A). (C) dI/dV mapping recorded at different energy positions (I = 500 pA). The quantum resonance state within each QC depends sensitively on the shape and size. Scanning parameters: 22 nm, -330 mV, 300 pA.

One of the three QCs in Figure S7A is not closed, with one Br atom missing. Interestingly, the QC with a missing Br atom also exhibits a bright domelike spot at negative bias and a dark hole at positive bias (dI/dV mapping), similar to the case of closed QCs. This suggests that quantum resonance states are also created in this QC. Fundamentally, this could be understood by the fact that the quantum states arise from the interference of scattered electron waves due to the comparable QC size and typical Fermi wavelength of Au(111) surface electrons, which are robust and one missing Br atom in the barrier has minimal effect on the confined states. This phenomenon has also been observed in other QCs, such as the system formed by covalently linked organic frameworks, where quantum resonance states were also observed in open QCs [8].

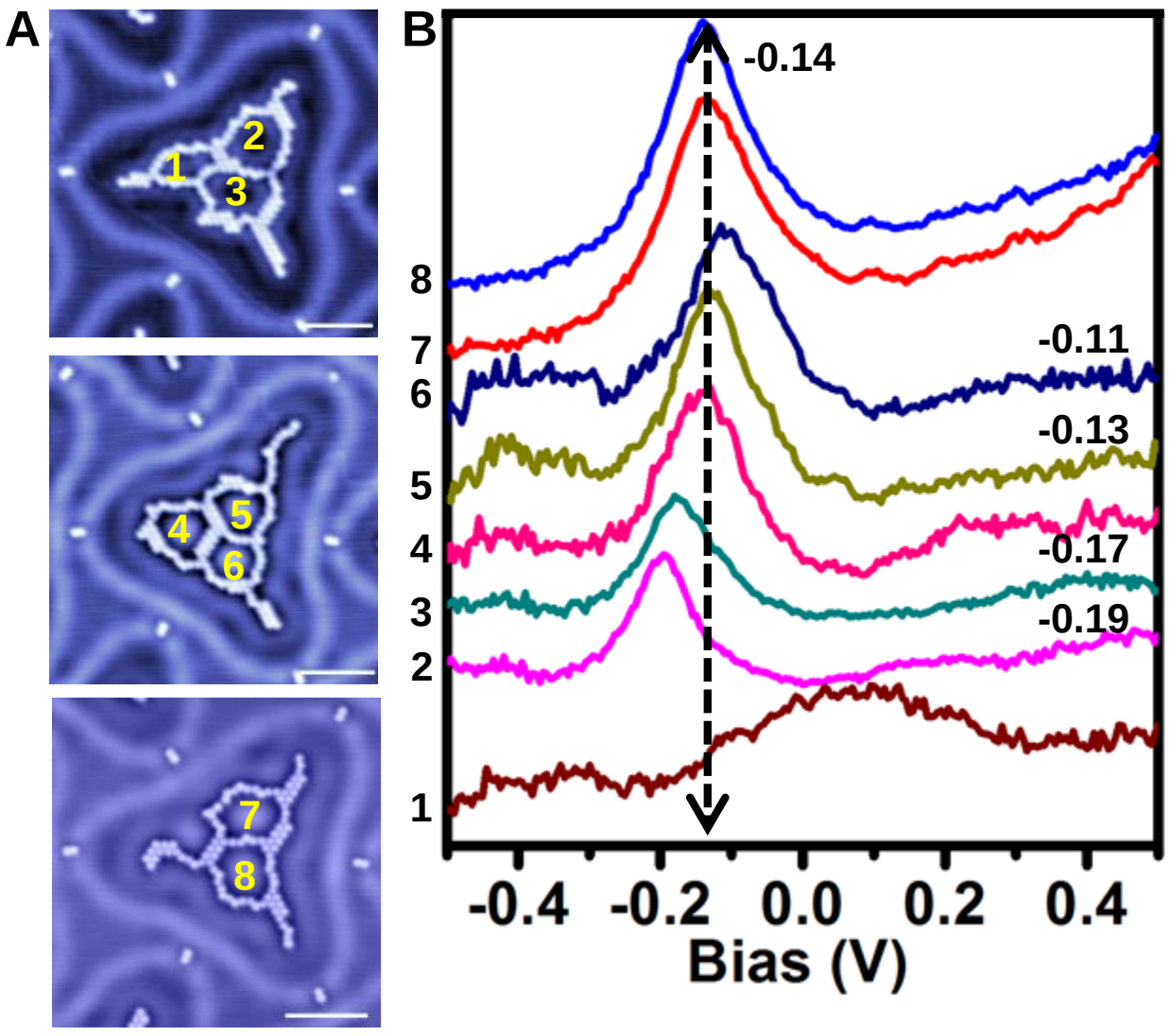

**Figure S8.** (A) STM images of QCs of different size (1~8). Scanning parameters (from top to bottom): 500 mV, 500 pA; -200 mV, 300 pA; 140 mV, 800 pA. Scale bar: 5 nm. (B) Point STS acquired at the center of the pores shown in (A).

The average diameters for pores 2~8 are measured to be 3.89, 3.85, 3.81, 3.81, 3.55, 3.81 and 3.81 nm, respectively. The peak positions of the point STS spectra acquired for pores 2~8 are -0.19, -0.17, -0.14, -0.13, -0.11, -0.14 and -0.14 eV, respectively (the size of pore 1 is too small for us to observe the QC states). Therefore, smaller pore size will generally lead to the shift of peak position to higher energy. This could be understood by the particle-in-box model: As the energy eigenvalue is determined by $E_{n,l} = \frac{Z_{n,l}^2}{2\mu R^2}$, where R is the radius of the nanopore, smaller nanopores will have higher energy eigenvalues (see more details in Figure S10).

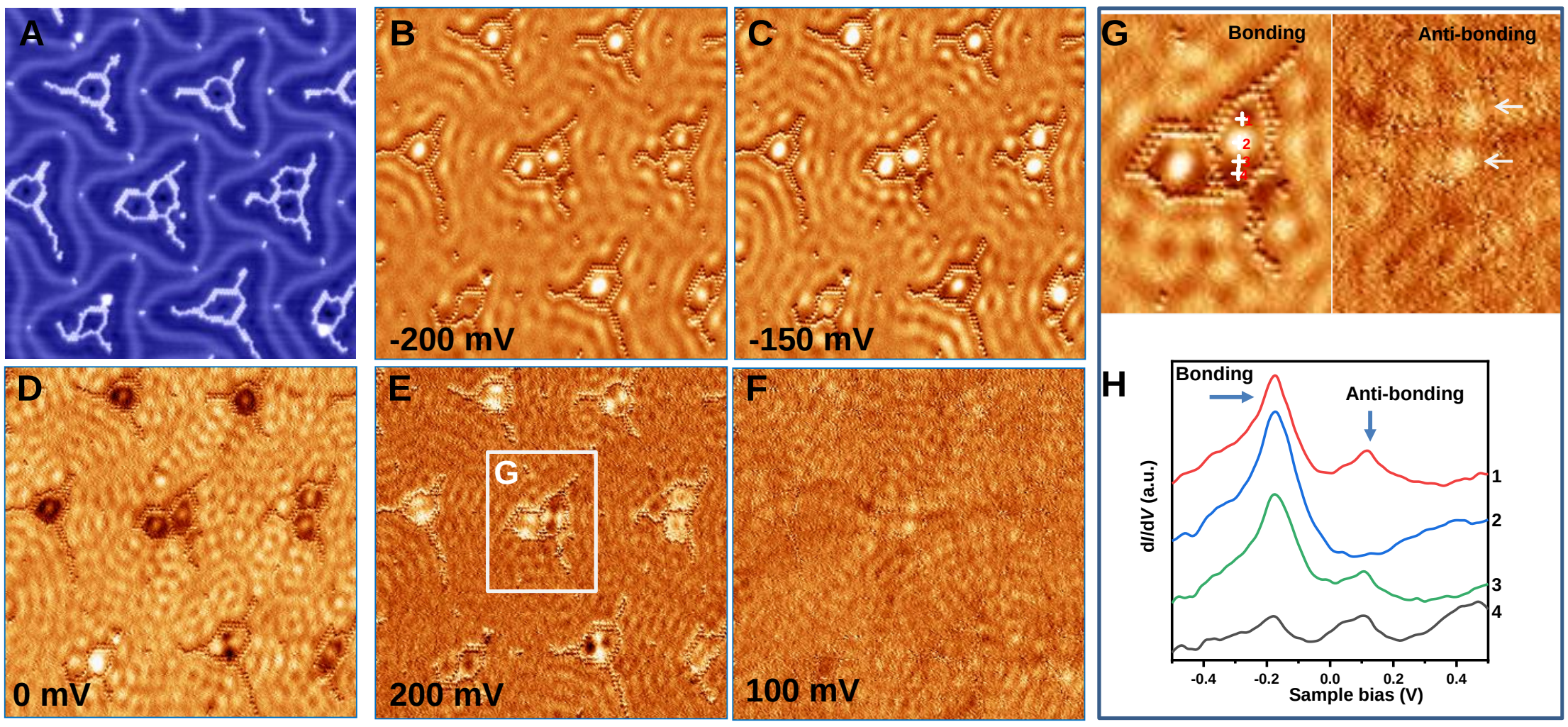

**Figure S9.** (A) STM image of various QCs with an average size of 3.7 nm. Scanning parameters: 48 nm, 250 mV, 300 pA. (B-F) Constant-current dI/dV maps recorded at different energy positions. The stretched QC is indicated by a white box in (E). (G) Constant-current dI/dV maps were recorded at energies of -0.2 and 0.1 V (I=500 pA),

corresponding to the bonding and anti-bonding states, respectively. (H) Point STS spectra taken at the position marked in panel (G).

As shown in the dI/dV maps in Figure S9, there are two bright spots and two dark holes that are very sensitive to the bias voltage in the dI/dV map within the stretched nanopore (Figure S9E-G). The length of the stretched nanopore is ~5.4 nm, so it can be viewed as two merged QCs with no inner barrier, which is similar to the case of stadium-shaped QC as discussed in the manuscript. The electronic states of this stretched QC derive from the hybridization of two P1 states that split into bonding and antibonding states. To show this, we performed point dI/dV measurements, as shown in Figure S9H. Clearly, an electronic state of ~-200 mV is observed around the center, and a pair of electronic states with energies of -200 and 100 mV can be found at the foci. This has the same characteristic with the stadium-shaped QC, where the bonding and antibonding states of $H_2$-like molecular orbitals are observed.

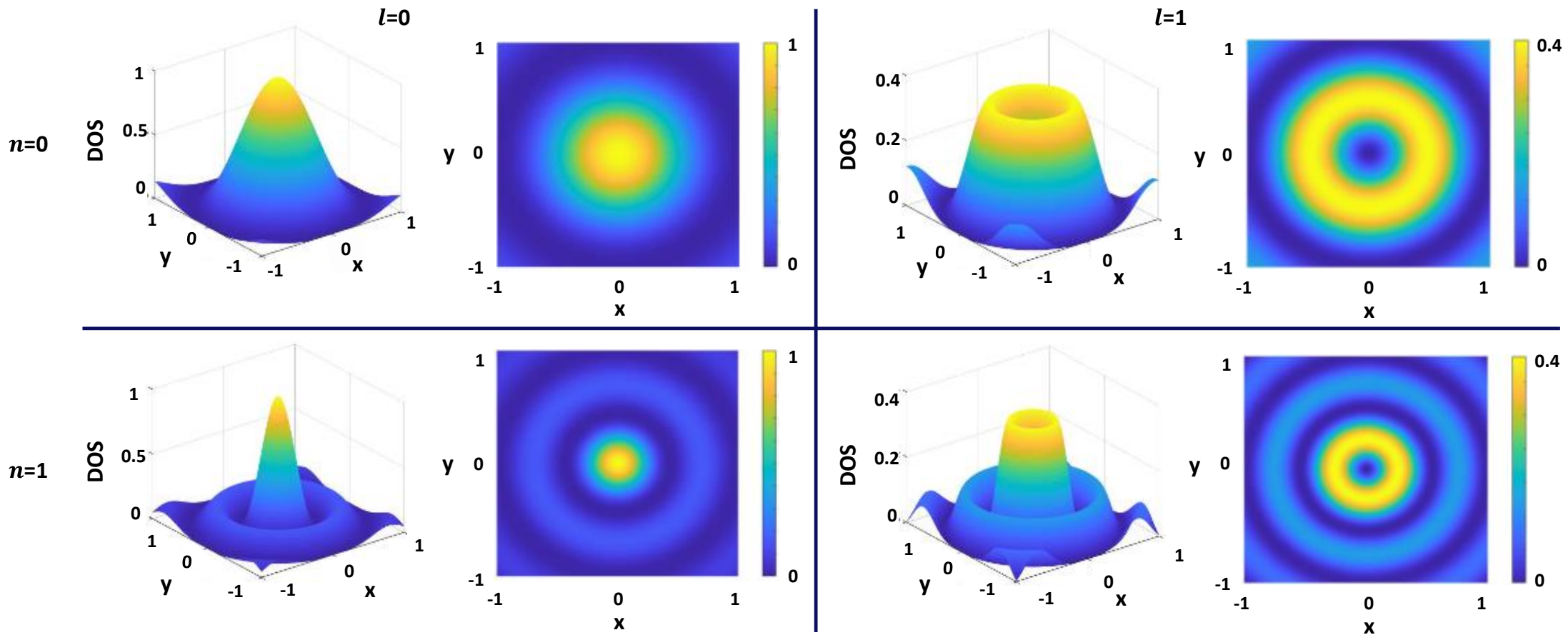

**Figure S10**. Particle probability distribution in a circular potential well. Here, the Schrödinger equation for a particle in a circular potential well can be written as, $\frac{1}{2\mu}\frac{d^2}{d^2r}\Psi(r) - \frac{1}{2r\mu}\frac{d}{dr}\Psi(r) + \left(\frac{1^2}{2\mu r^2}\right)\Psi(r) = E\Psi(r)$ , with $r = \sqrt{x^2 + y^2}$ . The eigenvalues and eigenstates are calculated to be $E_{n,l} = \frac{Z_{n,l}^2}{2\mu R^2}$ and $\Psi_{n,l} \propto J_l(Z_{n,l}, R)$ respectively. $n$ is the principle quantum number, $l$ is the angular momentum quantum number and $R$ is the radius of potential well.

Based on the particle-in-box model, the energy eigenvalues and eigenstates are $E_{n,l} = Z_{n,l}^2/2\mu R^2$ and $\Psi_{n,l} \propto J_l(Z_{n,l}, R)$ , respectively. Figure S10 shows the eigenstates and their distributions in real space obtained by solving the Schrödinger equation in a 2D infinite circular potential well. Clearly, the lowest energy level, $\Psi_{0,0}$ ($P_1$ state), has the highest DOS with states mainly distributed at the center of the ring. Meanwhile, as the eigenvalues, $E_{n,l}$, are inversely proportional to $R^2$, so smaller nanopore will have higher energy eigenvalue. Interestingly, our experimental results also support this, as discussed in Figure S8.

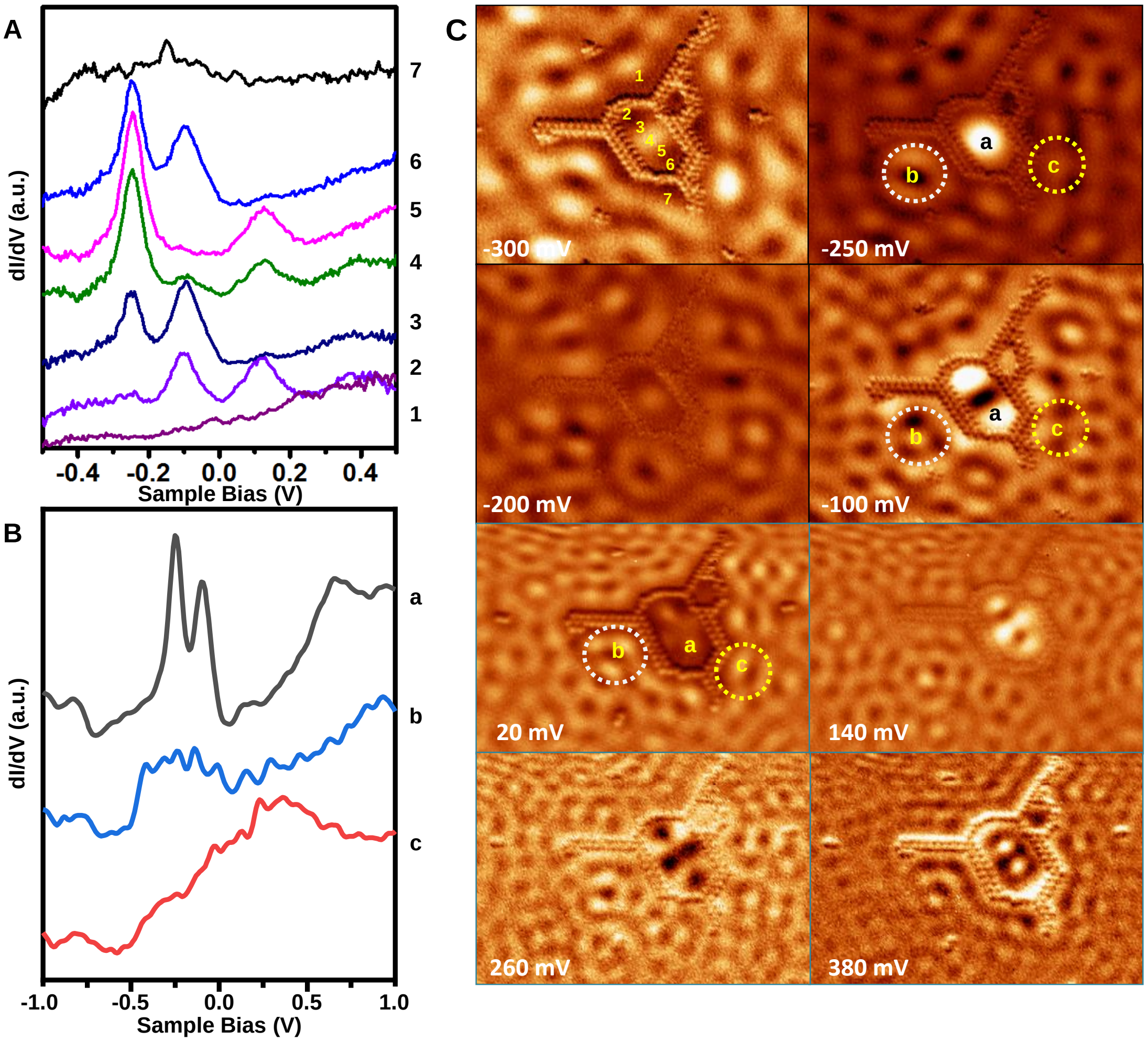

**Figure S11**. (A) Point STS acquired on different positions of the QC. Curves 1-7 correspond to the positions marked in (C). (B) Representative point STS spectra taken at the positions indicated in panel (C). (C) Constant-current dI/dV maps acquired under different energies. Scanning parameters: 22 nm, 420 mV, 500 pA.

The states observed outside the stadium-shaped QC arise from standing waves due to the scattering of surface state electrons with neighboring Br adatoms. Figures S11A and S11B shows the point spectra acquired inside and outside the stadium-shaped QC. A pair of electronic states were observed inside the QC (point a) at -0.25 and -0.1 V, corresponding to the bonding and anti-bonding states. In comparison, the spectra acquired on points 2 and 3 (outside the QC) are featureless, and the bonding and anti-bonding states cannot be observed. Here, points b and c are selected because of the contrast appearance in the dI/dV map, as highlighted by the dashed circles in Figure S11C.

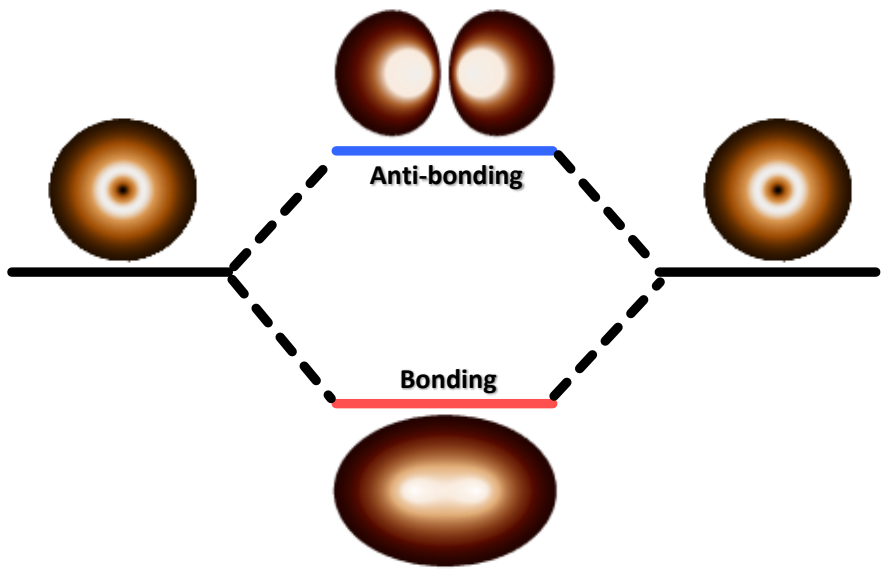

**Figure S12.** Calculated results for the hybridization of two atomic orbitals into $H_2^+$ molecular-like orbitals with bonding and antibonding states. The Schrödinger's equation for $H_2^+$ can be written as, $\left[-\frac{1}{2}\nabla^2 - \frac{1}{r_a} - \frac{1}{r_b} + \frac{1}{R_{AB}}\right]\psi = E\psi$, $R_{AB}$ is the distance of two nuclei A and B, $r_a$ ($r_b$) denotes the distance between electrons and nucleus A (B). The wave functions can be solved, with $\Psi_1 = \frac{1}{\sqrt{2+2S}}(\psi_a + \psi_b)$, $\Psi_2 = \frac{1}{\sqrt{2-2S}}(\psi_a - \psi_b)$, where $\psi_a = \frac{1}{\sqrt{\pi}}e^{-r_a}$, $\psi_b = \frac{1}{\sqrt{\pi}}e^{-r_b}$, $S = (1 + R_{AB} + \frac{R_{AB}{}^2}{3})e^{-R_{AB}}$. The spatial distribution can be determined by the probability $|\Psi|^2$.

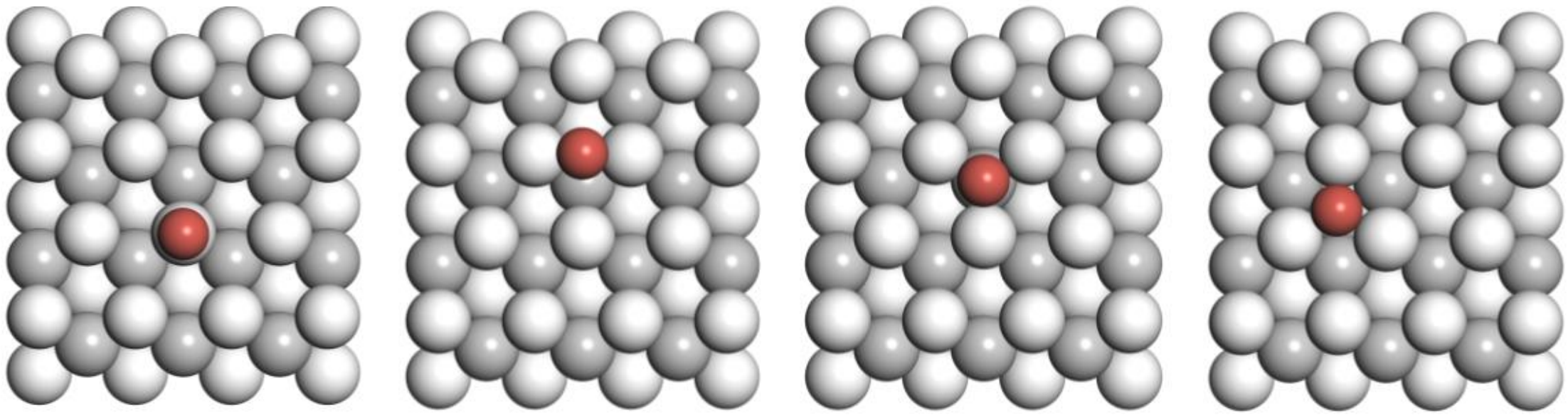

**Figure S13.** Different adsorption configurations of Br on different sites of Au(111) surface. From left to right: Top, bridge, hcp and fcc hollow sites.

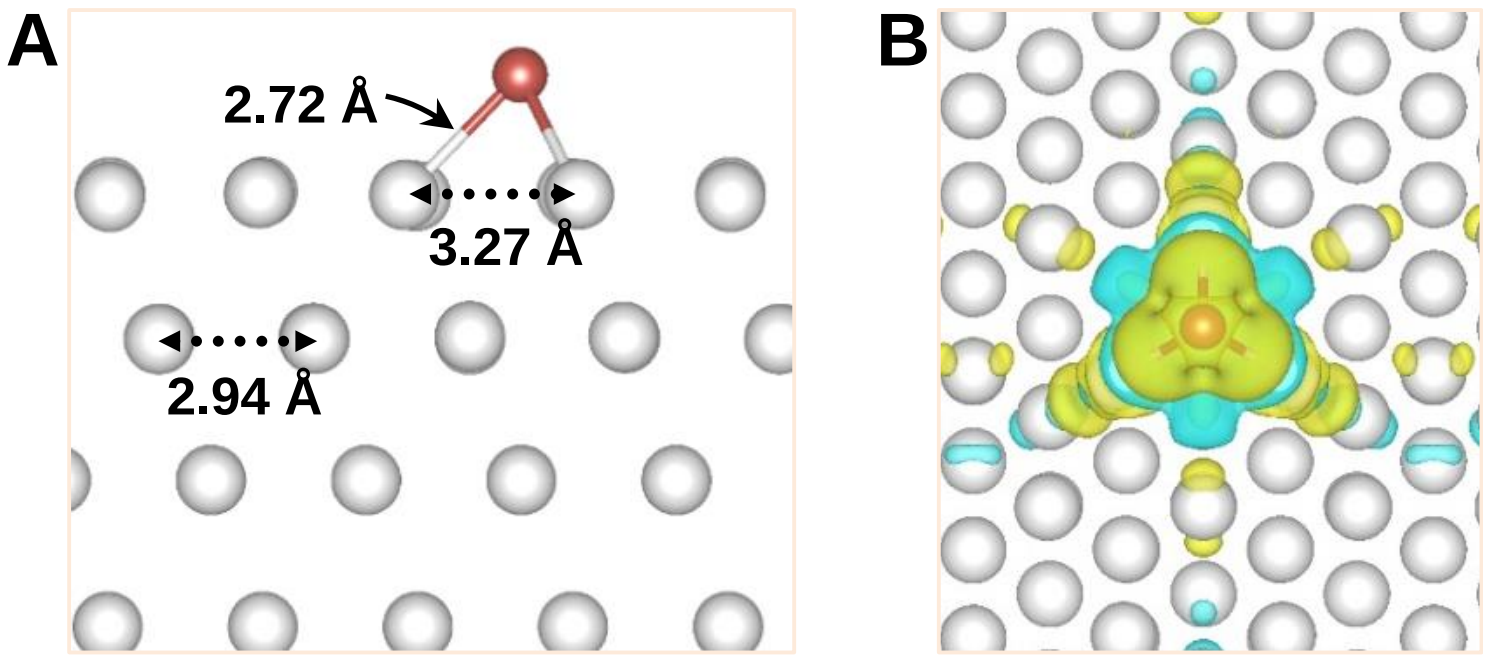

**Figure S14.** (A) Side view of Br atom adsorption at the fcc site of Au(111) with a bond length of 2.72 Å. The Au-Au bond length below Br extends to 3.27 Å. (B) Charge density redistribution upon adsorption of Br on Au(111) (isovalue= 0.005 $e$/Å$^3$).

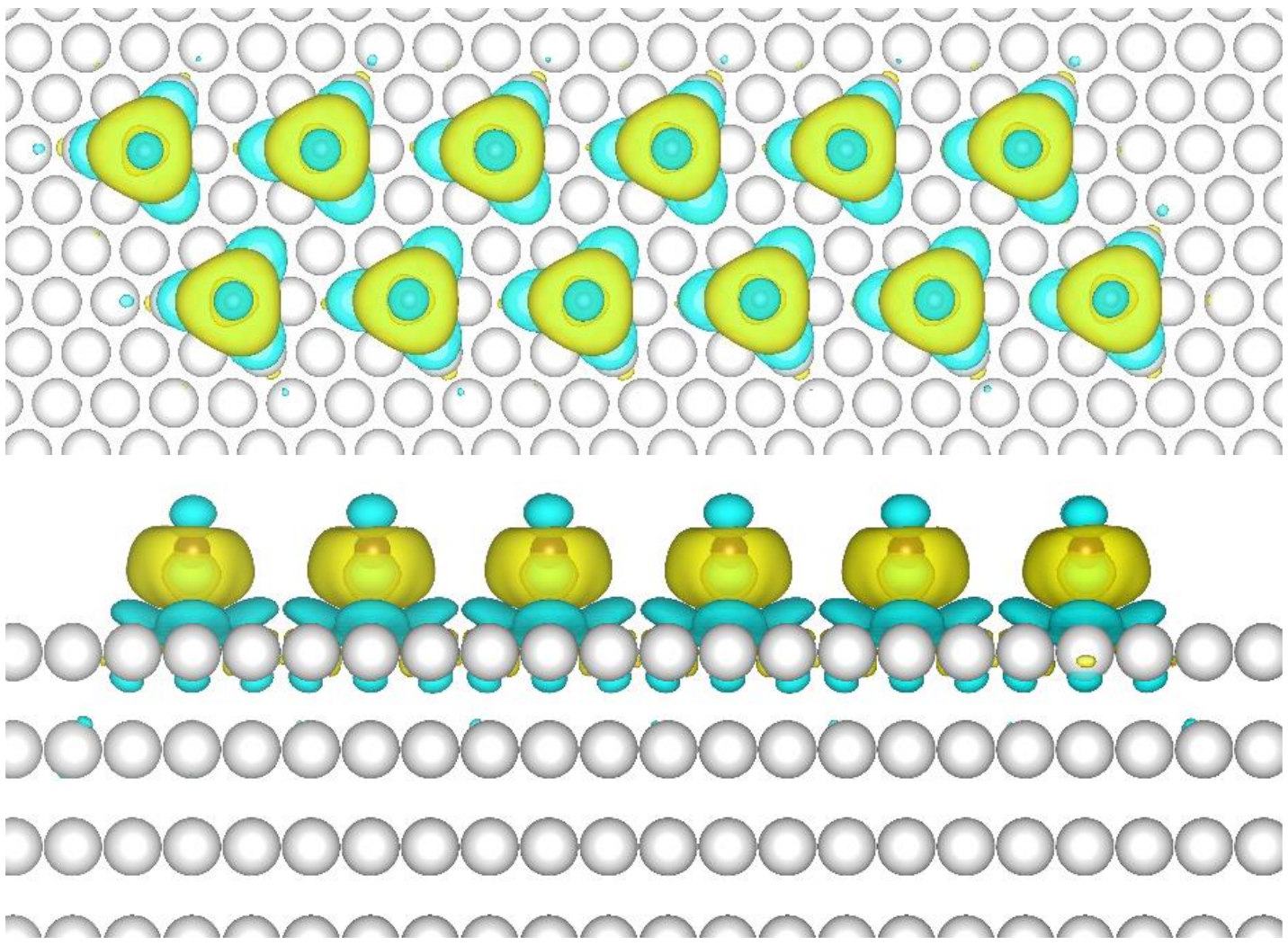

**Figure S15.** Top and side views of charge density redistribution for two parallel Br arrays along $[1\bar{1}0]$ direction (isovalue= 0.01 $e$/Å$^3$).

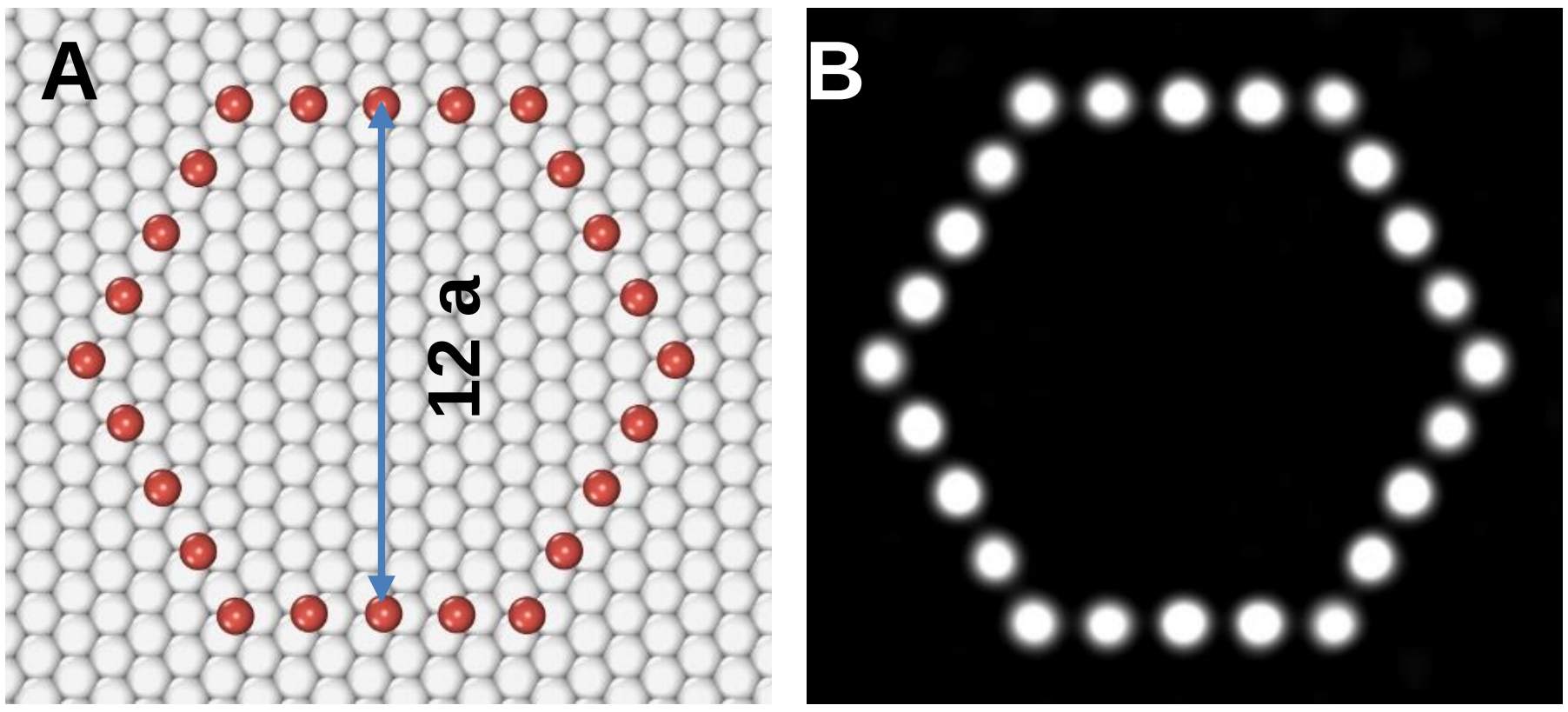

**Figure S16.** (A) Atomic structure and (B) simulated STM image of a hexagonal QC.

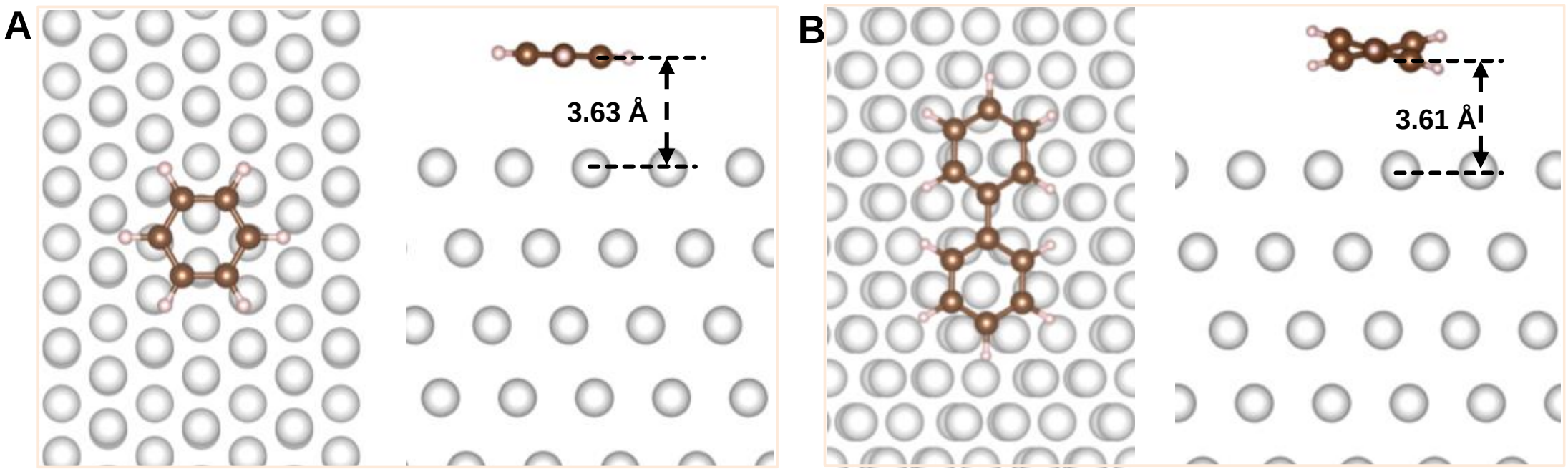

**Figure S17.** (A) Top and side views of the optimized structure of benzene adsorption on Au(111) surface. (B) Same as (A) for biphenyl adsorption on Au(111). The adsorption distance is indicated for both molecules.

We performed DFT calculations on the structure and energetic properties of representative carbon structures on Au(111), as shown in Figure S17. Complete dehalogenation of $C_6Br_6$ results in the formation of a $C_6$ ring [9]. Carbon radicals are highly active and can couple with each other, which can also be passivated by hydrogen in the vacuum chamber [10]. Here, we have considered benzene and biphenyl molecules on Au(111) surface. With the inclusion of van der Waals interaction, benzene adopts a flat-lying configuration on the surface with an adsorption distance of 3.63 Å and an adsorption energy of -1.14 eV. For biphenyl, the adsorbed molecule is nonplanar. The calculated adsorption distance is 3.61 Å and the adsorption energy is -2.15 eV. The large adsorption distances of both benzene and biphenyl molecules suggest facile diffusion of these structures, so that these carbon species can be aggregated into large clusters and islands. This is in accordance with our STM observations, for which carbon islands with large size over several tens of nanometers can be observed (see STM images in Figure S2).

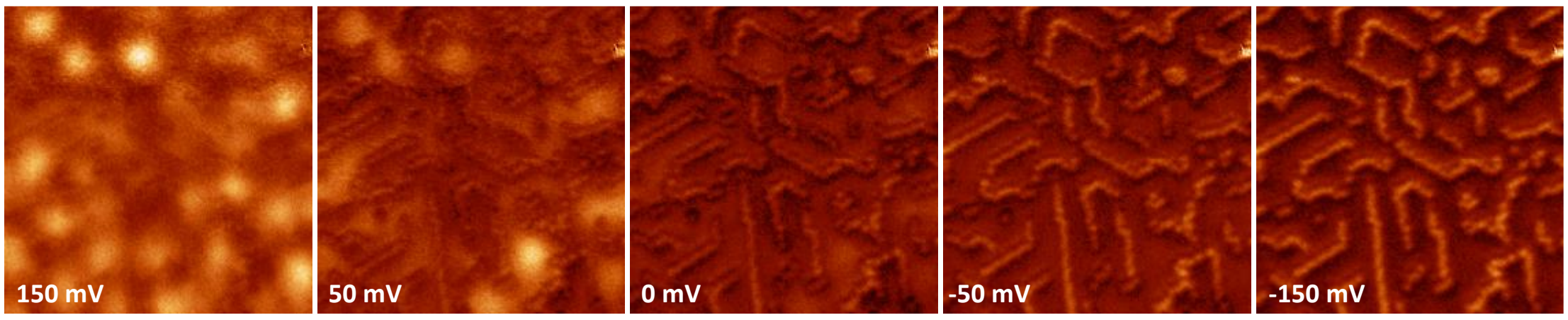

**Figure S18.** dI/dV mapping acquired on Br adatom adsorption on Ag(111). The Br adatoms aggregate into short chains that are randomly distributed on the surface.

**Table S1.** Calculated adsorption energies ($E_{ad}$, in eV) for Br adsorption on different sites of Au(111) surface.

| Sites | top | bridge | hcp | fcc |
|---|---|---|---|---|
| $E_{ad}$ | -1.09 | -1.36 | -1.34 | -1.36 |

## REFERENCES

1. Horcas, I.; Fernández, R.; Gómez-Rodríguez, J. M.; Colchero, J.; Gómez-Herrero, J.; Baro, A. M. WSXM: A Software for Scanning Probe Microscopy and a Tool for Nanotechnology. *Rev. Sci. Instrum.* **2007,** *78*, 013705.

2. Kresse, G.; Furthmüller, J. Efficiency of Ab-Initio Total Energy Calculations for Metals and Semiconductors Using a Plane-Wave Basis Set. *Comput. Mater. Sci.* **1996,** *6*, 15-50.

3. Kresse, G.; Furthmüller, J. Efficient Iterative Schemes for Ab Initio Total-Energy Calculations Using a Plane-Wave Basis Set. *Phys. Rev. B* **1996,** *54*, 11169-11186.

4. Perdew, J. P.; Burke, K.; Ernzerhof, M. Generalized Gradient Approximation Made Simple. *Phys. Rev. Lett.* **1996,** *77*, 3865-3868.

5. Grimme, S.; Antony, J.; Ehrlich, S.; Krieg, H. A Consistent and Accurate Ab Initio Parametrization of Density Functional Dispersion Correction (DFT-D) for the 94 Elements H-Pu. *J. Chem. Phys.* **2010,** *132*, 154104.

6. Tersoff, J.; Hamann, D. R. Theory and Application for the Scanning Tunneling Microscope. *Phys. Rev. Lett.* **1983,** *50*, 1998-2001.

7. Nanayakkara, S. U.; Sykes, E. C. H.; Fernández-Torres, L. C.; Blake, M. M.; Weiss, P. S. Long-Range Electronic Interactions at a High Temperature: Bromine Adatom Islands on Cu(111). *Phys. Rev. Lett.* **2007,** *98*, 206108.

8. Peng, X.; Mahalingam, H.; Dong, S.; Mutombo, P.; Su, J.; Telychko, M.; Song, S.; Lyu, P.; Ng, P. W.; Wu, J.; Jelínek, P.; Chi, C.; Rodin, A.; Lu, J. Visualizing Designer Quantum States in Stable Macrocycle Quantum Corrals. *Nat. Commun.* **2021,** *12*, 5895.

9. Gao, W.; Kang, F.; Qiu, X.; Yi, Z.; Shang, L.; Liu, M.; Qiu, X.; Luo, Y.; Xu, W. On-Surface Debromination of $C_6Br_6$: $C_6$ Ring Versus $C_6$ Chain. *ACS Nano* **2022**, *16*, 6578.

10. Zang, Y.; Jiang, T.; Gong, Y.; Guan, Z.; Liu, C.; Liao, M.; Zhu, K.; Li, Z.; Wang, L.; Li, W.; Song, C.; Zhang, D.; Xu, Y.; He, K.; Ma, X.; Zhang, S.-C.; Xue, Q.-K. Realizing an Epitaxial Decorated Stanene with an Insulating Bandgap. *Adv. Funct. Mater.* **2018,** *28*, 1802723.